\documentclass[sigconf,nonacm,balance=false]{acmart}
\usepackage{popets}

\usepackage{amsmath}
\usepackage{array}
\usepackage{colortbl}
\usepackage{framed}
\usepackage{tikz}
\usetikzlibrary{positioning,arrows.meta,shadows.blur}
\usepackage{pgfplots}
\pgfplotsset{compat=1.18}

\definecolor{cbzInk}{RGB}{16,60,96}        
\definecolor{cbzCert}{RGB}{18,122,84}      
\definecolor{cbzFalse}{RGB}{196,72,52}     
\definecolor{cbzTint}{RGB}{225,235,245}    
\definecolor{cbzRow}{RGB}{238,244,250}     
\definecolor{cbzGuess}{RGB}{190,110,20}    
\definecolor{cbzModel}{RGB}{110,110,110}   
\definecolor{cbzData}{RGB}{252,240,205}    
\definecolor{cbzNN}{RGB}{214,228,245}      

\newcommand{\prov}[1]{}   

\newcommand{\hl}{\rowcolor{cbzRow}}

\begin{document}

\title[Reading Is Not Leaking]{Reading Is Not Leaking: Local, Auditable Measurement and Reduction of Inference Exposure from Public Footprints}

\author{Mahmudul Faisal Al Ameen}
\affiliation{%
  \institution{Independent researcher}
  \country{France}}
\email{faisal@mahmudulfaisal.com}

\renewcommand{\shortauthors}{Faisal Al Ameen}

\begin{abstract}
Anyone with a public footprint leaks facts that were never stated, and language models
make the inference cheap. We present a framework for measuring and reducing this
inference exposure that runs on the owner's own CPU with no language model at analysis
time, instantiated on organisations and on individuals. It starts from a measurement
result: scoring an inference system against the target's private truth conflates how
well the system reads the record with how much the record leaks. On a 128-question
instrument over sixteen synthetic firms, almost half of the questions are never
answered correctly by any of six readers, four of them language models, and a
majority-class guess accounts for most of every reader's score. We therefore separate
reading accuracy from leakage rate and introduce an injection protocol that creates
cells with known support. Our analyser combines rules, statistical solvers and a
106M-parameter encoder trained from scratch that marks verbatim evidence and never
generates text; every answer carries a graded certificate whose recorded proof
replays. Its certified answers are correct in 93\% of resolved cases, against 49--73\%
for the language models' quote-backed answers, whose citations are produced alongside
the answer rather than deriving it; with plain-prose articles in the record, 70\% of
its evidence-bearing answers rest on evidence that establishes them, against 18--56\%
for the models. A constrained defence that rewrites each fact's carrier as a true but
coarser statement hides every single-carrier fact from four language-model adversaries
at 40\% lower edit cost than deletion. On sixteen synthetic people the guessing term
is larger still, and a decoy planner with no language model halves the correct answers
of the estimator it targets without transferring to a second.

\end{abstract}


\maketitle

\section{Introduction}
\label{sec:intro}

A public footprint is written by many hands for many purposes, and no one reads it
as a whole. For a person it is posts, likes, reviews, check-ins and the traces kept
in platform accounts; for an organisation it is registry filings, product pages, job
postings, conference talks and customer case studies. An adversary can read all of
it. Large language models infer personal attributes from users' posts cheaply and at
scale~\cite{staab2024beyond}, and competitors, acquirers, short sellers and state
actors apply the same capability to organisations, inferring a pending acquisition,
a supplier dependency, a hiring freeze or the runway left before the next raise.

The owner of a footprint needs two things: \emph{what can be inferred about me}, and
\emph{what would change if I changed the record}. Answering both on the owner's own
machine matters twice over. The material is sensitive by definition, so sending it
to a hosted model to learn how much it leaks defeats the purpose; and a tool that
needs a data-centre GPU is out of reach for most individuals and small
organisations. We therefore build an analyser that runs on an ordinary CPU without a
language model at analysis time, and make every answer carry verbatim evidence.

Both needs depend on a benchmark that measures inference honestly, and this paper
starts from a finding we did not expect. The standard way of scoring such systems,
comparing each answer with the target's private truth, cannot separate a system that
reads the record well from a record that leaks; on our instrument it mostly ranks
how systems guess. Almost half of the questions are never answered correctly by any
of six readers, four of them language models, and a majority-class guess alone
accounts for most of every reader's score. The remedy is to measure reading and
leakage separately and to create cells whose support is known.

\paragraph{One framework, two instantiations.}
Nothing in the measurement, the analyser's contracts or the defence loop depends on
the kind of entity being read. What changes between domains is a \emph{domain pack}:
the fact schema, the question bank, the readers and rules, the sources and the
ledger of admissible interventions (\S\ref{sec:architecture}). We evaluate the
framework on organisations (\S\ref{sec:org}) and on individuals
(\S\ref{sec:personal}), where our work began; the second pack was added without
changing a single answer or certificate of the first.

\paragraph{Contributions.}
\begin{enumerate}\setlength{\itemsep}{1pt}
\item \textbf{A measurement result and a metric.} Scoring against private truth
mixes leakage, reading and guessing. We define \emph{reading accuracy} on supported
cells and \emph{leakage rate} as separate quantities, and give an injection protocol
that localises, per question, where a stated fact is lost
(\S\ref{sec:measurement}). On sixteen synthetic firms, 56 of 128 questions are
never recovered by any reader and a majority-class guess scores 479 of 1,842 cells
(\S\ref{sec:unsupported}).
\item \textbf{A local, auditable, domain-general analyser.} Rules, statistical
solvers and dedicated readers over cited evidence, with no language model at
analysis time and an encoder of our own design trained from scratch; every answer
carries a graded certificate whose proof replays, and guesses are labelled as
guesses (\S\ref{sec:architecture}).
\item \textbf{Measured comparisons.} Against a rule extractor on held-out real
companies (\S\ref{sec:extractioneval}); against four language models on the same
instrument, prompt and scorer (\S\ref{sec:comparison}), where we distinguish
\emph{forward derivation} of evidence from \emph{post-hoc attribution}; and on a
real company (\S\ref{sec:realfirm}).
\item \textbf{A second instantiation.} The same core on sixteen synthetic people
in public and private-export conditions: a stronger guessing term, evidence reading
that gains on public footprints only, and deletion and decoy defences whose effect on
statistical inferences does not transfer to a second estimator
(\S\ref{sec:personal}). Decoys are split into a deterministic \emph{planner}, which
decides the signal and is measured here, and a \emph{renderer}, where a language
model can later write the content.
\item \textbf{A constrained defence.} Interventions on the public record that
contain no false or misleading statement of material fact, planned against the
transparent analyser and evaluated for transfer to language-model adversaries
(\S\ref{sec:defence}).
\end{enumerate}

\paragraph{Scope of claims.}
The sixteen organisations are synthetic development fixtures, read many times while
the system was built; the extractor comparison uses real company text with held-out
firms; the real-company case reports coverage, not accuracy; the personal results are
development measurements on synthetic people whose cohorts informed the rules. These limits are stated where each number
appears and collected in \S\ref{sec:limits}, and Appendix~\ref{app:provenance} maps
every quoted figure to the artefact that produces it.

\section{Problem and threat model}
\label{sec:problem}

\paragraph{Targets and footprints.}
A \emph{target} is an entity with a private state, a person or an organisation, and
a \emph{public footprint}: the documents and records about it that anyone can read
up to a cutoff date. Some items are written by the target, others by third parties
about it.

\paragraph{Adversary.}
The adversary reads the footprint up to the cutoff and answers questions about the
target's private state. It has no access to non-public data but unlimited reading
capacity. We consider two adversary classes: a \emph{transparent analyser}, ours,
whose every answer can be traced to the evidence that produced it
(\S\ref{sec:architecture}); and \emph{language-model readers} given the entire
footprint in context (\S\ref{sec:comparison}).

\paragraph{Instrument.}
Exposure is measured against a fixed \emph{question bank}: the questions an
adversary would want answered, each with a typed answer (yes/no, category, band,
date window, name, word, sentence) and, where a question only makes sense given
another, a \emph{gate} on its parent's answer (a deal's counterparty is asked only
if a deal exists). A \emph{key} gives the private truth for each scorable (target,
question) cell, and a type-aware \emph{comparator} decides equivalence, treating
numeric bands, units and ranges explicitly and leaving ambiguous comparisons
\emph{unresolved} rather than wrong. The key describes the target's private truth,
not what its record shows; that distinction is the subject of
\S\ref{sec:measurement}. The bank belongs to the domain pack. For organisations it
covers ownership, finances, customers and suppliers, strategy, hiring, technology,
incidents and pending transactions (\S\ref{sec:org}); for individuals, identity,
location and routine, health, beliefs, relationships and finances
(\S\ref{sec:personal}).

\paragraph{Defender and constraint.}
The defender is the footprint's owner, acting on the part of the footprint it
controls. A defence may not contain a false or misleading statement of material
fact. Interventions are classed by a \emph{ledger}: generalisation and timing shift
are admissible; suppression is admissible only with a stated reason why the absence
will not itself be read as a signal~\cite{dye1985,jung1988}; additional
non-material signals need sign-off; misleading material statements are
excluded~\cite{grossman1981,milgrom1981}. Which actions are available is
domain-specific: an organisation can reword its own pages and filings within
disclosure law, while an individual can typically delete, edit or restrict the
visibility of their own posts but cannot rewrite what others have published.

\section{Reading is not leaking}
\label{sec:measurement}

\subsection{The conflation}
Let $C$ be the set of scorable cells (target, question) and $t(c)$ the key's
truth. A reader $R$ answers $a_R(c)$ or abstains, and the usual score is
$\mathrm{Acc}(R)=|\{c: a_R(c)\equiv t(c)\}|/|C|$. Let $S\subseteq C$ be the cells
whose answer is \emph{supported} by the public record: stated in it, or entailed by
it under the question's definition. Then
\begin{equation}
\mathrm{Acc}(R)\;=\;L\cdot\mathrm{RA}(R)\;+\;G(R),
\label{eq:split}
\end{equation}
where $L=|S|/|C|$ is the \emph{leakage rate} of the record, a property of the
record alone; $\mathrm{RA}(R)=|\{c\in S: a_R(c)\equiv t(c)\}|/|S|$ is the
\emph{reading accuracy} of the reader on supported cells; and
$G(R)=|\{c\notin S: a_R(c)\equiv t(c)\}|/|C|$ is \emph{lucky guessing}. The third
term rewards answering unsupported cells, and a reader that abstains where the
record is silent, the behaviour a defender wants, is penalised relative to one that
guesses the base rate. Scoring against private truth therefore measures $L$,
$\mathrm{RA}$ and guessing policy together. The decomposition holds for any target
and question bank; how large the guessing term is depends on how predictive base
rates are in the domain, which is itself worth reporting.

\subsection{What we report}
For every reader and every defence we aim to report:
\begin{itemize}\setlength{\itemsep}{1pt}
\item \textbf{reading accuracy} on supported cells, with abstentions counted as
misses and base-rate answers excluded by construction;
\item the \textbf{leakage rate} of the record, independent of any reader (in this
paper only bracketed, \S\ref{sec:unsupported});
\item the \textbf{precision of evidence-backed answers} with their coverage, since
these are the answers a user acts on;
\item the private-truth score, for comparability with prior work, always next to the
base-rate reference.
\end{itemize}
Labelling $S$ directly requires a human adjudicator per cell. Until that labelling
exists, two measurements bracket it: a pooled-reader upper bound on the unsupported
part (\S\ref{sec:unsupported}), and cells whose support is known by construction,
which the next protocol provides.

\subsection{Creating supported cells: the injection protocol}
\label{sec:injectionmethod}
Supported cells are hard to label at scale, but they can be \emph{created}. Take
questions that no reader recovers, write documents in which the target's own public
record states the true answer plainly, add them to a copy of the footprint, and
follow each fact through the pipeline. Because support is known by construction,
every loss can be attributed to a stage. Each (target, question) pair is classified
as \emph{recovered}, \emph{extracted but no solver answers}, \emph{answered
wrongly}, \emph{answered but unresolved by the comparator}, or \emph{not
extracted}. The same injected footprints serve as secrets for the defence
(\S\ref{sec:defence}), since each fact's carriers are known. Documents written with
the key in view make this a ceiling probe on plainly stated facts, not an estimate
of real-world leakage.

\section{Architecture}
\label{sec:architecture}

\subsection{Design principles}
Six commitments shape every component.
\begin{enumerate}\setlength{\itemsep}{1pt}
\item \textbf{Local and accessible.} Analysis and defence run on the owner's own
machine. No component sends the target's material to a third party, and every step
runs on an ordinary laptop CPU; the tool presupposes neither a data-centre GPU nor a
paid model subscription.
\item \textbf{No language model at analysis time.} The system combines
deterministic code and statistics with one neural model of our own design, an
encoder (\S\ref{sec:extraction}). A generative model is reserved for rendering,
for example turning a decoy plan into post or search text (\S\ref{sec:decoy}); it
is gated by the deterministic components, which decide what signal to produce, rather
than trusted to decide it.
\item \textbf{Cited evidence.} Every answer carries its support: evidence rows with
verbatim spans, the clause that produced the answer and any derivation.
\item \textbf{Guesses are labelled.} A base-rate answer is never shown as a
finding; abstention is a first-class outcome.
\item \textbf{Measured reliability.} Every learned or rule-based component earns
its place against the majority-class reference on data it was not tuned on;
components that do not earn it remain reported arms but do not answer.
\item \textbf{Domain packs.} Everything that knows what kind of entity is being
read sits in a replaceable pack over a domain-general core
(\S\ref{sec:domainpacks}).
\end{enumerate}
Table~\ref{tab:example} in \S\ref{sec:org} follows one real record through every
stage described below, showing what each stage's data becomes.

\begin{figure*}[t]
\centering
\small
\begin{tikzpicture}[
  node distance=4mm and 6mm,
  box/.style={draw=cbzInk!70, line width=0.5pt, rounded corners=3pt, align=center,
              minimum height=10mm, text width=24mm, inner sep=2pt, font=\footnotesize,
              blur shadow={shadow blur steps=4, shadow xshift=0.4pt, shadow yshift=-0.6pt, shadow opacity=12}},
  det/.style={box, fill=white},
  nn/.style={box, fill=cbzNN, draw=cbzInk},
  store/.style={box, fill=cbzData, draw=cbzGuess!70, text width=21mm},
  arr/.style={-{Stealth[length=2mm]}, thick, cbzInk!80},
  lab/.style={font=\scriptsize\itshape, text=cbzModel}]
\node[store] (rec) {public footprint\\{\scriptsize documents, records,}\\{\scriptsize activity, prose}};
\node[det, right=of rec] (ing) {ingest\\{\scriptsize documents, passages,}\\{\scriptsize republication groups}};
\node[nn, right=of ing] (ext) {extract\\{\scriptsize rules \emph{+} own encoder}};
\node[store, right=of ext] (ev) {evidence rows\\{\scriptsize key, value, subject,}\\{\scriptsize assertion, span}};
\node[det, right=of ev] (sol) {solve\\{\scriptsize readers, rules, stats,}\\{\scriptsize labelled prior}};
\node[det, below=7mm of sol] (fus) {fuse\\{\scriptsize gates, abstention,}\\{\scriptsize certificate L0--L3}};
\node[store, left=of fus] (ans) {answers\\{\scriptsize + why-view}\\{\scriptsize (rules, patterns)}};
\node[det, left=of ans] (leak) {leak map\\{\scriptsize carriers, minimal}\\{\scriptsize hitting sets}};
\node[det, left=of leak] (plan) {plan\\{\scriptsize target, ledger gate,}\\{\scriptsize search}};
\node[det, left=of plan] (int) {intervene\\{\scriptsize generalise, delay,}\\{\scriptsize suppress}};
\draw[arr] (rec) -- (ing);
\draw[arr] (ing) -- (ext);
\draw[arr] (ext) -- (ev);
\draw[arr] (ev) -- (sol);
\draw[arr] (sol) -- (fus);
\draw[arr] (fus) -- (ans);
\draw[arr] (ans) -- (leak);
\draw[arr] (leak) -- (plan);
\draw[arr] (plan) -- (int);
\draw[arr] (int.north) -- node[lab, left] {changes to the record} (rec.south);
\draw[arr, dashed] (plan.north) to[bend left=12] node[lab, sloped, above, pos=0.42, yshift=-1pt] {analyser in the loop} (ext.south);
\node[lab, above=1mm of ext] {\textcolor{cbzInk}{own neural model}};
\node[lab, left=1mm of rec, anchor=east, text=cbzInk] {analysis};
\node[lab, left=1mm of int, anchor=east, text=cbzInk] {defence};
\end{tikzpicture}
\caption{The analyser (top row) and the defence loop (bottom row), identical for every
domain. White: deterministic code and statistics; blue: our encoder; amber: data. The
planner re-runs the analyser on every candidate change, so the defence is optimised
against the same transparent reader that produced the findings; its effect on other
adversaries is measured separately (\S\ref{sec:defence}).}
\Description{Block diagram: a top row from public footprint through ingest, extract, evidence rows and solve, and a bottom row from fuse through answers, leak map, plan and intervene, with an arrow back to the footprint.}
\label{fig:architecture}
\end{figure*}
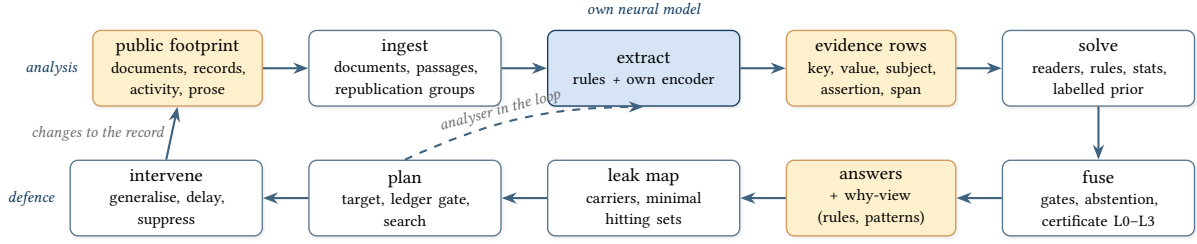

\subsection{Domain-general core and domain packs}
\label{sec:domainpacks}
The pipeline of Figure~\ref{fig:architecture}, its contracts, the runtime guard,
fusion and certification, the encoder's architecture and training code, the
measurement of \S\ref{sec:measurement} and the defence loop are shared by every
domain. A domain pack supplies the rest (Table~\ref{tab:domainpacks}). Adding a
domain means writing a pack and training the encoder's task heads on its schema,
not changing the pipeline.

\begin{table*}[t]
\centering\small
\caption{What the core shares and what each domain pack supplies.}
\label{tab:domainpacks}
\begin{tabular}{@{}>{\raggedright}p{0.10\textwidth}>{\raggedright}p{0.28\textwidth}>{\raggedright}p{0.26\textwidth}>{\raggedright\arraybackslash}p{0.27\textwidth}@{}}
\toprule
\rowcolor{cbzTint}
layer & domain-general core & organisation pack (\S\ref{sec:org}) & personal pack (\S\ref{sec:personal})\\
\midrule
sources & document and passage contracts, character offsets, dates with confidence, cutoff, republication groups & registries, company pages, filings, job boards, press & platform data exports, posts, likes, searches, reviews, location history\\
evidence & row contract: key, typed value, subject, assertion status, verbatim span, stable row identifier & 22 business fact types; subject is the target or another party & personal fact types (health, beliefs, relationships, location, routine, finances); subject is self, a named other or a public figure\\
extraction & rules engine; encoder architecture, value grammar, training code & encoder fine-tuned on company pages and filings & rules over text and native fields (21 observation keys, subject and register fields); encoder adaptation to informal text planned\\
solvers & readers, Horn rules with proofs, aggregates, change points, latent state, classifiers; routing against the majority reference & 128-question readers; 42 rules; 30 business patterns & readers for 32 questions, statistical solvers for 10 more over activity logs; 39 of 97 questions covered\\
answers & fusion, gates, abstention, certificates L0--L3, display policy & --- & ---\\
instrument & typed questions with gates, frozen key, comparator, split metric, injection protocol & 128 questions, 16 synthetic firms & 97 active questions, 16 synthetic personas in five cohorts (public and private-export)\\
defence & leak map, hitting sets, planner, ledger gate, transfer test & ledger from disclosure economics; generalise, suppress & own content only; delete, redact or restrict visibility, with receipts\\
training data & family-split generation, licensed teacher, held-out entities & firm simulator and factory; web pages and SEC filings & synthetic personas; real data only in a consented study, never sent to a hosted model\\
\bottomrule
\end{tabular}
\end{table*}

\subsection{Contracts}
Components communicate only through typed records, which is what lets a rule, a
statistical solver and a neural model be exchanged behind one interface.
\begin{itemize}\setlength{\itemsep}{1pt}
\item \textbf{Query}: target, cutoff date, task (whole window, state now, or a
forecast with a horizon), and a time axis that is synthetic (months 1--18) or
calendar-dated for real targets.
\item \textbf{Document and passage}: a source snapshot with retrieval time,
publication time when known, source kind, and a republication group so that copies
of one announcement count as one witness; passages carry exact character offsets
into their document.
\item \textbf{Evidence row}: key, typed value, subject, assertion status (observed,
asserted, denied, planned, hypothetical, unknown), dates, extractor identity and
confidence, a verbatim span validated against its passage, and a stable identifier
(a hash of document, offsets, key, value, assertion and subject).
\item \textbf{Candidate}: an answer to one question, a score, a basis (direct,
inference, prior, abstention), the rows its clause consumed, the passages it
consulted, stated assumptions and any arithmetic.
\end{itemize}

\subsection{Separation of training and analysis}
A runtime guard admits into the analysis window only declared, hash-verified code
and fitted assets; label files, private state and the evaluation key stay outside
it, so no answer can be influenced by the truth it is scored against. Everything
learned is fitted beforehand on data separated by \emph{family}: related entities (a
firm and its rival, a person and their household) are never split across training,
calibration and test. In the organisation pack, training data comes from three
sources: a structured simulator (5,000 labelled firms in seconds, used to fit
statistical solvers and calibrate rules); a family-split factory with public records
and prose (1,000 rival families, split 700/150/150); and, for the encoder, real
company text labelled by an open-weight teacher. The sixteen authored firms of the
benchmark are development fixtures and are never used for fitting.

\subsection{Ingest and extraction}
\label{sec:extraction}
Registries, platform exports, feeds and crawled pages become documents; prose
becomes passages with exact offsets; boilerplate is dropped; relevance to the target
is scored per passage; and republications are grouped. For real targets every date
is a calendar date with a confidence, and cutoffs reject anything published after
the query date.

Structured feeds are extracted exactly by field rules. Prose passes through two
extractors whose rows share one schema: rules for numbers, money, percentages,
dates and fixed phrasings, and our encoder for everything that depends on context:
whose fact a statement concerns, and whether it is asserted, denied, planned or
hypothetical. The organisation pack's vocabulary has 22 business fact types
(organisation mentions, releases, announcements, money, percentages, headcount,
people and roles, locations, customer segments, supplier dependencies, stated
objectives, business changes, certifications, technology and others) and two row
types anchored on the instrument's own questions, a \emph{relation} (``this span
states the answer to question X'', with yes/no carried by the assertion) and a
\emph{metric} (``this span states the number for question X''), each consumed by
at least one solver.

\paragraph{Encoder.}
A 106M-parameter transformer encoder (12 layers, width 768, 12 heads, 1,024-token
passages; 97M parameters in the encoder and 8.4M in task heads) with a 16k-piece
byte-pair tokenizer~\cite{sennrich2016bpe} trained by our own code. Its layers are
standard published components~\cite{vaswani2017attention,su2024roformer}; the task
architecture is ours.
\begin{itemize}\setlength{\itemsep}{1pt}
\item \emph{Head-anchored rows.} Each row is anchored on the mention that identifies
it (the organisation's name, the number, the person) and tagged per key with
begin/inside labels. Anchoring on the head rather than the sentence separates
several facts stated in one sentence; in a labelled sample of 3,943 passages, 23\%
of organisation rows share their sentence with another organisation row.
\item \emph{Row heads.} From the anchor, classifiers predict subject, assertion
status and closed-set fields; pointer heads~\cite{vinyals2015pointer} select the
supporting span and each text field, with a learned null slot for absent fields.
\item \emph{Value grammar.} Numbers, scales, currencies and dates are parsed by
deterministic rules from the mention a pointer lands on; the network never generates
a value. Every output string is a verbatim substring of the input, so a fabricated
value is structurally impossible.
\end{itemize}

\paragraph{Training.}
Masked-token pretraining from random initialisation~\cite{devlin2019bert} under a
fixed seven-hour budget on one GPU, following~\citet{geiping2023cramming}: 6.35B
tokens seen from a 3.0B-token corpus of English Wikipedia (CC~BY-SA),
FineWeb-Edu~\cite{penedo2024fineweb} (ODC-By) and our own crawl of training
companies' public pages, with the held-out firms excluded even from unlabelled
text. Fine-tuning uses evidence rows written by teachers whose licence or terms of
service permit training on their outputs (DeepSeek-V3.2, MIT~\cite{deepseek2025v32};
GLM-5.2, MIT; OpenAI's Codex); a row is kept only if its span occurs verbatim in the
passage. The fine-tuning text is passages from company websites and SEC 10-K filings
(\S\ref{sec:lessons} explains why the filings are needed) and, for the
question-anchored rows, free-form articles about generated firms; training uses
per-key loss weighting and is validated on 14 companies disjoint from training. The
architecture,
tokenizer training and both training stages are domain-general: a new pack reuses
the pretrained encoder, continues pretraining on its own register of text where that
differs, and trains new task heads for its schema. Extraction quality on held-out
real companies is reported in \S\ref{sec:extractioneval}; the encoder processes a
passage in about 0.3\,s on a laptop CPU (1,858 passages in 507\,s).

\subsection{Solvers}
Each solver maps evidence to candidate answers for the question types it declares.
\begin{itemize}\setlength{\itemsep}{1pt}
\item \textbf{Dedicated readers}, one per question family, each declaring which
evidence it consumes, how it chooses among candidates (explicit ranking, nearest
stated date, scope cues), which assertion statuses may answer (a plan never answers
a factual question; a denial answers only questions that ask about absence), and
when it abstains (conflicting values).
\item \textbf{Aggregates}: explicit formulas over rows (counts, sums, shares).
\item \textbf{Relations}: Horn rules with proofs~\cite{ullman1988databases},
combined by maximum rather than noisy-OR unless independence is stated.
\item \textbf{Timing and bursts}: Poisson change-point and scan
statistics~\cite{hinkley1970changepoint,kulldorff1997scan}, answering the event
window rather than the first public mention.
\item \textbf{Latent state}: a hidden Markov model for hiring
stance~\cite{rabiner1989hmm}, emitted per question only where calibration shows it
beats the majority class.
\item \textbf{Reference arms}: per-question classifiers (logistic regression,
boosted trees) and a majority-class prior, fitted on the simulator. They are
reported for comparison; on real firms they refuse to answer, because they are
fitted only on synthetic whole-window questions.
\end{itemize}
A question is routed to a solver only if that solver beats the majority reference
on calibration families. On 300 reserved generated firms, this per-question
selection (which keeps the majority reference for 46 questions) gains 1.3 correct
answers per firm over the prior; on the sixteen authored firms the same selection
does not beat the prior, a transfer failure from generated to authored records that
the measurement of \S\ref{sec:measurement} helps to explain.
\prov{countermodel/out/experiments/improvement-20260922/REPORT.md}

\subsection{Fusion, gates and certificates}
\label{sec:certificates}
Fusion combines candidates using each solver's measured reliability, applies the
instrument's gates (a dependent question is answered only if its parent's answer
opens it, and a guessed parent never closes a gate silently), and abstains when
candidates conflict or when the best score is below the display threshold. Each
answer is graded by the strength of its support: \emph{L3}, exact rows and the
clause that fired, whose re-execution on those rows alone reproduces the answer;
\emph{L2}, exact rows not yet certified; \emph{L1}, a pointer to the relevant
passage, or a statistical association; and \emph{L0}, a labelled base-rate answer.
Weak links are graded rather than discarded: a pointer to the right passage still
helps a reviewer. Because rows carry stable identifiers, a certificate names the
exact rows its clause consumed rather than the passages it consulted, and
de-duplicated copies of a row stay listed as its members.

L3 certifies \emph{sufficiency}: the recorded clause reproduces the answer from its
recorded rows. It does not certify that the answer is true, and it is distinct from
\emph{necessity}: an answer can survive the removal of its support because a second,
independent proof exists or because the base-rate answer coincides with it. We
record these as \emph{redundant} and \emph{prior-coincident} support. They are not
defects, and a redundant fact is exactly what a defence must cover twice
(\S\ref{sec:defence}). A citation is \emph{decorative} only if it is insufficient.

\subsection{Explanation layer}
\label{sec:explanation}
Beyond answering, the analyser explains \emph{why}, without generating text. In the
organisation pack a \emph{rulebook} of 42 hand-written business rules (for example,
job postings that stop after a period of hiring indicate a freeze) runs in the
Horn-rule engine with proofs down to verbatim spans. Each rule's precision is
measured on held-apart generated firms and the rule receives one of four statuses.
An \emph{earned} rule, whose lower confidence bound beats the majority reference on
the firms where it fires, may answer; one rule is earned, at 0.74 precision. An
\emph{informative} rule, whose lower bound exceeds its answer's base rate but not the
prior, is shown as a \emph{weak guess} with its premises and measured precision,
never as an answer; three rules are informative, and the one we added for this
purpose (postings that continue at a low, even rate indicate steady hiring)
replicates on sealed test families at 0.73 precision against a base rate of 0.47.
A \emph{hypothesis} has no measured lift, and its premises are displayed as context
rather than support; an \emph{uncalibrated} rule has premises that never occur in
generated firms. The distinction matters because a premise displayed next to a
guess without measured lift would be exactly the post-hoc attribution we criticise
in language-model readers (\S\ref{sec:comparison}). A \emph{pattern layer} detects
30 business patterns (restructuring, covenant pressure, release cadence, hiring
ahead of capacity, executive media before announcements) with calibrated detectors,
attributes each detection to the rows whose removal lowers it most, and renders a
templated sentence citing them; 8 of the 30 earned on calibration families and all 8
replicated on reserved test families.
\prov{PROGRESS.md sections B1 and B2; out/experiments/weak-links-20260924/}

\subsection{Deployment: the pipeline as a certified program}
\label{sec:contract}
The whole analyser is a Python package with one neural asset (the encoder, 404\,MB)
and small fitted tables, and it runs on a laptop CPU. In the desktop application the
pipeline is written as a program in Kimiya, an orchestration language of our own
whose runs produce a certificate recording what ran and what was checked
(Figure~\ref{fig:contract}). Each stage of Figure~\ref{fig:architecture} is one call,
bound to a deterministic Python function, and the program checks properties between
stages rather than trusting a single opaque call: that the public record was read and
produced documents and evidence; that no candidate cites a record the analyser could
not see, and that only the prior may answer while citing nothing; that inputs and
fitted assets are byte-identical to what the opening stage declared; and, at the end,
that no language model took part and that every answer is grounded. No agent is
declared anywhere in the program, so the certificate's egress record is empty, and
its network check reports what an audit hook observed during the run rather than what
any package states about itself. The contract is therefore the definition of the
analysis: the trace records the pipeline that actually ran, and a reviewer reads the
guarantees in the program instead of inferring them from code. The encoder is also
exposed as a local endpoint, so that the same language can call it as a model agent
that proposes evidence, with certainty left to the deterministic checks that follow
it.

\begin{figure}[t]
\scriptsize
\definecolor{shadecolor}{named}{cbzRow}
\setlength{\FrameSep}{3pt}
\begin{shaded*}
\vspace{-2pt}
\begin{verbatim}
sid := open_(inp)       -- declare cohort, instrument,
                        -- assets; arm the guards
rec := records(sid)     -- 1 public record
check field(rec, "records") > 0
corpus := ingest(sid)   -- 2 documents, passages
check field(corpus, "documents") > 0
ev := extract(sid)      -- 3 typed evidence rows
check field(ev, "rows") > 0

r := solve(sid, "reader,aggregate,graph,
                 dynamics,latent")  -- 4 evidence
c := classify(sid)      --   fitted classifiers
p := priors(sid)        --   population prior
check attributed(sid)   -- nothing cites an unseen
                        -- record; only a prior may
                        -- answer citing nothing
a_rules := fuse(sid, "rules")     -- 5 one answer
a_class := fuse(sid, "classify")  --   per question,
a_prior := fuse(sid, "prior")     --   per arm
routed    := route(sid, "false")  -- 6 product answer
selective := route(sid, "true")   --   provable only
check unchanged(sid)    -- inputs and assets match
                        -- the declared hashes
res := close(sid)       -- 7
check no_llm(res)
check grounded(res)
\end{verbatim}
\vspace{-6pt}
\end{shaded*}
\caption{The analysis pipeline as a Kimiya contract, from the desktop application
(abridged: input loading and progress messages omitted). Every stage is a deterministic
Python function; no agent or model is declared, so the certificate's egress record is
empty, and every \texttt{check} is recorded in the run's certificate.}
\Description{Listing of the pipeline contract: seven stages with checks between them.}
\label{fig:contract}
\end{figure}

\section{Case study: organisations}
\label{sec:org}

This section instantiates the framework on organisations. The organisation pack
was built first because a firm's footprint is large, mostly written by the firm
itself and public by design, so real-company text can be collected and labelled
without touching personal data.

\subsection{Instrument}
The question bank has 128 questions in fifteen areas (ownership and control,
finances, customers and suppliers, strategy, hiring, technology, incidents, pending
transactions and others), 41 of them gated on a parent answer. The questions are
posed about sixteen synthetic firms in eight rival pairs, each with a complete
private state and a generated public record over an 18-month window. The frozen key
has 1,842 scorable cells (of 2,048), produced from each firm's private specification
by deterministic skeleton anchors and a model panel under a verification contract.
The ledger of admissible interventions follows the disclosure-economics analysis
summarised in \S\ref{sec:problem}.

\subsection{A worked example: one record through the pipeline}
\label{sec:example}
Table~\ref{tab:example} follows a single record of firm F05A, a fintech, through
every stage of Figure~\ref{fig:architecture}, using the artefacts the saved run
produced rather than a schematic. The same firm supplies two contrasts, a labelled
guess and a certified error, and the defence step on its injected copy.

\begin{table*}[t]
\centering\small
\caption{One record of firm F05A through the pipeline. Every value is copied from the
saved run (\S\ref{app:provenance}); row identifiers are truncated. Offsets are
half-open character positions in the document.}
\label{tab:example}
\begin{tabular}{@{}>{\raggedright\bfseries}p{0.13\textwidth}>{\raggedright\arraybackslash}p{0.84\textwidth}@{}}
\toprule
\rowcolor{cbzTint}
stage & what the data becomes\\
\midrule
record (input) &
\texttt{F05A-0025}, channel A12 (executive interview), date M11.14, kind
\texttt{interview\_transcript}. A podcast transcript with the CFO. Excerpt of one
answer: ``\ldots\ volumes have grown every quarter, and we're managing the business
toward sustainable margins, not growth at any cost. On capital, any growth-stage
fintech is thinking about its balance sheet and options all the time, and that's
true for us too. I don't have anything to announce there.''\\[3pt]
ingest &
Document \texttt{F05A-0025}: 5,885 characters, source kind external (native kind
interview transcript), republication
group \texttt{origin:F05A-0025}. Five overlapping passages with exact offsets
(\texttt{p0-1534}, \texttt{p1418-2820}, \ldots), each scored 1.0 for relevance to
the target.\\[3pt]
extract &
18 evidence rows from this document. Three of them: \newline
(a) encoder: key \texttt{objective\_posture}, value \texttt{\{posture: growth,
primary: true\}}, subject F05A, assertion \texttt{asserted}, span ``we're managing
the business toward sustainable margins, not growth at any cost.'', offsets
1261--1340, \texttt{row\_id 8ea82de8}, extractor \texttt{extractor\_v3\_encoder}
(checkpoint hash recorded). \newline
(b) encoder: key \texttt{org\_mention}, value \texttt{\{name: Stripe, context:
competitor, primary: false\}}, subject empty (the row is about Stripe, not the
target), span ``pry enterprise accounts
away from the likes of Stripe and Adyen'', offsets 2756--2819. \newline
(c) rules: key \texttt{announcement}, value \texttt{\{verb: launch, \ldots\}} from the
token ``launching'' at 2661--2670, subject empty, assertion \emph{unknown}: a row no
reader will consume, because its subject is not established.\\[3pt]
solve &
Question UB-27b, ``What is its primary objective posture?'', receives two candidates.
The \emph{reader} fires clause \texttt{reader.UB-27b.objective\_posture} on row (a)
alone: answer \texttt{growth}, basis direct, score 0.8, proof \texttt{\{consumed:
[8ea82de8], context: [], kind: reader\}}. The \emph{prior} proposes \texttt{growth}
with an empty proof (kind prior).\\[3pt]
fuse &
Answer \texttt{growth}, solver reader, score 0.8. Decision: winner reader; competitor
prior, grade L0, ``lower selection score''. Certificate: grade \textbf{L3}, support
\texttt{[8ea82de8]}, sufficiency \texttt{passed} (re-running the clause on that one
row reproduces \texttt{growth}), one quote with document, offsets and verbatim
span.\\[3pt]
score (outside the analyst) &
Key truth \texttt{growth}: correct. The quote is shown to the reader, who may note
that ``sustainable margins, not growth at any cost'' could also be read as a
profitability posture; the certificate guarantees that this row produced this answer,
not that the reading is beyond dispute.\\[3pt]
contrast: a labelled guess &
Question UE-18a, ``Is a funding round being prepared?'': no evidence candidate. The
prior answers \texttt{N} with score 0.81, grade \textbf{L0}, reason ``population
prior; pointers are not evidence for this prediction''. Key truth \texttt{Y}: the
guess is wrong. The CFO's ``thinking about its balance sheet and options \ldots\
nothing to announce'' is exactly the silence that no reader may turn into evidence
(\S\ref{sec:comparison}).\\[3pt]
contrast: a certified error &
Question UG-01a, ``What is its current headcount?'': the aggregate clause
\texttt{latest\_headcount} consumes the row ``Revised salary letters are going out to
around 400 of the team'' and answers \texttt{400}; grade L3, the proof replays. Key
truth \texttt{450}. The row was read correctly and the formula answered a different
question; the certificate localises the defect to one clause.\\[3pt]
defence (injected copy) &
Secret UA-08a, ``Who holds final approval over major strategic and investment
decisions?'', carried by one
sentence of the injected article \texttt{F05A-0091}: ``Final approval over major
strategic and investment decisions rests with Board of directors.'' Generalise
rewrites it to ``\ldots\ rests with its senior governance.'' The analyser no longer
recovers the secret, and neither does any of the four language-model adversaries
(\S\ref{sec:defence}).\\
\bottomrule
\end{tabular}
\prov{out/experiments/provenance-offsets-20260924/capture/{stores,predictions}/original/F05A*; cohorts/F05A.json; decoy188 cohorts/generalise/F05A.json}
\end{table*}

\subsection{Extraction on held-out real companies}
\label{sec:extractioneval}
We compare the encoder with the rule extractor on identical passages and labels
(Table~\ref{tab:extraction}). The labels are rows written by a strong commercial
model and used only for evaluation; the reference is therefore agreement with a
careful annotator, not ground truth, and two strong annotators themselves recover
only 57--78\% of each other's rows while agreeing on 94.7\% of assertion statuses
and on every subject where they match.

\begin{table}[t]
\centering\small
\caption{Extraction on held-out real companies (Apple and Palantir; 291 passages
never used in training): same passages, labels and matcher (same key and overlapping
span). Encoder at threshold 0.5 unless noted.}
\label{tab:extraction}
\resizebox{\columnwidth}{!}{%
\begin{tabular}{@{}lcccccc@{}}
\toprule
\rowcolor{cbzTint}
 & F1 & recall & precision & subject & assertion & value\\
\midrule
rule extractor & 0.291 & 0.225 & 0.412 & 0.439 & 0.404 & \textbf{0.905}\\
\hl our encoder & \textbf{0.402} & \textbf{0.435} & 0.374 & \textbf{0.765} & 0.736 & 0.734\\
\hl our encoder, threshold 0.7 & 0.399 & 0.340 & \textbf{0.484} & \textbf{0.765} & \textbf{0.749} & 0.801\\
\bottomrule
\end{tabular}}
\prov{ai-libs/out/extractor_eval/pod_v3/eval_v3_heldout_opus.json; ai-libs/out/extractor_eval/heldout_opus_rules.json}
\end{table}

On firms never seen in training the encoder recovers about twice the rows of the
rule extractor and is about twice as accurate on subject and on assertion status;
the rules remain better on numeric values, so the analyser keeps both. On validation
companies the encoder scores F1 0.387 (0.399 at threshold 0.7), and its six
additive keys reach recall of 0.68 (headquarters location), 0.54 (stated
objective), 0.31 (business change), 0.27 (supplier dependency; precision 0.67 at
threshold 0.7), 0.33 (hiring focus) and 0.17 (customer segment).

\subsection{Where the score comes from}
\label{sec:unsupported}
Supported cells are not labelled directly. A conservative upper bound on the
unsupported part comes from asking whether \emph{any} reader ever recovers a
question. We pooled six readers on the sixteen firms: our evidence-only solvers,
with and without the encoder, and four language models given the whole public
record in context under an identical prompt (\S\ref{sec:comparison}).
Table~\ref{tab:recovery} gives the result.

\begin{table}[t]
\centering\small
\caption{Recovery of the 128 questions across sixteen firms by the pooled readers.}
\label{tab:recovery}
\resizebox{\columnwidth}{!}{%
\begin{tabular}{@{}lr@{}}
\toprule
\rowcolor{cbzTint}
questions never answered correctly for any firm & 56 (44\%)\\
\quad of which never even attempted by any reader & 19\\
questions recovered on 1--3 firms & 27\\
questions recovered on 8 or more firms & 26\\
questions recovered on all 16 firms & 1\\
\midrule
cells correct, majority-class guess (no evidence) & 479 / 1,842\\
cells correct, evidence-only arm & 50 / 1,842\\
\bottomrule
\end{tabular}}
\prov{ai-libs/out/unanswerable_items.json (56 / 19 re-checked on the headcount-fix-20260925 capture); headcount-fix-20260925/comparison-table/table.json}
\end{table}

The 19 never-attempted questions read as a list of what firms do not publish: the
ultimate beneficial owner and its share, majority voting control, market share,
total debt, an acquisition target and its signing window, an IPO window, the share
of revenue from the largest product line, months of cash runway, the largest
near-term obligation, the share of the workforce affected by a reduction, the
direction of paid-acquisition spend, the number of active enterprise opportunities,
the time to qualify an alternative supplier, which licence, which class of legal
claim and which class of regulator. This is the point of the instrument, which asks
what an adversary would want, and it is why private-truth scoring is dominated by
cells no reader can support.

Two cautions apply. A question can be recoverable yet missed by all six readers,
so 56 is an upper bound; adding the latest reader lowered it from 61 to 56 and the
never-attempted count from 31 to 19. That the count moves with the reader pool is
itself the argument of this section: any figure of this kind, and any single
reader's score, is partly a statement about the readers, which is why leakage must
be measured on the record rather than inferred from readers' failures. And for the
37 questions attempted but always answered wrongly, the comparator's strictness
matters; the injection protocol separates those cases.

\paragraph{Anatomy of a guess-dominated score.}
Of the combined analyser's 512 correct answers, 42 are certified from evidence
(L3), 5 rest on a statistical association (L1) and 465 are the labelled base-rate
guess. A backward search over the extracted rows, restricted to rows relevant to
each question, finds the following. For 359 of the 465, the question is a yes/no or
sentence question and the guess answers ``no'': is an acquisition pending, is a
funding round being prepared, is there a regulatory investigation. For most firms
the truth is no, and a firm never publishes that it is \emph{not} being acquired, so
no evidence can exist; these answers score as correct and nothing leaked. A further
68 are state judgements (legal form, security maturity, headcount trajectory,
pricing power, hiring stance, profitability, transaction posture) whose premises are
mostly already extracted, in 67 of the 86 such cells we examined, but which no rule
with measured lift connects to the answer; the weak-link rules of
\S\ref{sec:explanation} are the response, and the calibration data rather than the
rules is currently the limit (\S\ref{sec:lessons}). For 34 no relevant row carries
the value, and for 4 relevant evidence exists that no reader consumes. Silence,
in other words, is the largest single source of correct answers on this benchmark,
which is why our analyser labels it as a guess and why the readers'
behaviour on exactly these cells is examined in \S\ref{sec:comparison}.
\prov{out/experiments/support-check-relevant-20260924/, premise-gap-20260924/}

\subsection{Where a stated fact is lost}
\label{sec:injection}
We applied the protocol of \S\ref{sec:injectionmethod} to twenty questions: ten
that no reader ever attempted and ten that readers attempted but always answered
wrongly. For each firm, every question with a real truth value (not gate-closed) is
stated once, paraphrased, by sentence templates filled with the firm's truth from
the key, and grouped into two to four news articles of 162--251 words per firm in
the register of the firm's existing prose: 235 (firm, question) pairs in 63
articles. Table~\ref{tab:injection} follows the 235 facts through the pipeline
before and after dedicated readers were added.

\begin{table}[t]
\centering\small
\caption{The 235 injected facts before and after dedicated readers. Before injection,
none was recovered. $^{\dagger}$Before: no key can carry the fact. After: ten
cells whose support the reader did not assess and two without a typed relationship.}
\label{tab:injection}
\resizebox{\columnwidth}{!}{%
\begin{tabular}{@{}lrr@{}}
\toprule
\rowcolor{cbzTint}
 & before readers & after\\
\midrule
reached an evidence row & 193 (82\%) & ---\\
\quad no solver claims the question & 151 (64\%) & 0\\
\quad answered, wrong value & 20 & 2\\
\quad answered, comparator unresolved & 7 & 33\\
no usable evidence row$^{\dagger}$ & 42 & 12\\
\hl \textbf{recovered and correct} & \textbf{15 (6\%)} & \textbf{188 (80\%)}\\
\midrule
recovered, rules only & 5 & 188\\
recovered, rules + our encoder & 15 & 188\\
\bottomrule
\end{tabular}}
\prov{inject20-20260923T111332Z/summary.json; inject20-readersv2-20260923T141418Z/summary.json}
\end{table}

On these template articles reading was not the bottleneck: four of five stated
facts became evidence and then had nowhere to go. The protocol localised the loss per question. Seven questions had
no consumer at all, three had no key able to carry the fact (an affirmed or denied
proposition such as ``a viable substitute exists''), and two picked the wrong
number from a passage containing several. These counts became the specification for
dedicated readers, after which 188 of 235 facts were recovered with two wrong
answers. Questions that readers had previously attempted and always answered
wrongly recovered at the same rate once the fact was present (8 of 147 against 7 of
88 never-attempted questions, before readers), so they were never harder to read;
they had been answered from the wrong evidence. Of the 33 unresolved cells most are
the same value in another rendering (\texttt{USD 1e+07-2.5e+07} against
``\$10--25M''), a comparator issue rather than a reading one.

Before dedicated readers, the encoder tripled recovery (5 to 15), entirely through
money and percentage rows. After them, rules alone recover 188 and the encoder adds
nothing, because the injected articles are written in regular phrasing that rules
parse. The position reverses on real text (\S\ref{sec:realfirm}): the same rules
produce none of the relation rows the readers consume on a real company's pages, and
the encoder is what feeds them.

\paragraph{Unseen phrasing (pre-registered).}
The 188 is a development figure: the readers were specified from this probe's own
losses, and the articles' sentence frames were written on the same side as the
rules. We therefore froze the analyser (code, encoder, key; 167 files hashed) before
a second set of articles existed and had an independent model write the same 235
facts in eight document types the templates never used (interviews, podcasts,
analyst notes; 2\% six-word overlap with the templates), confirmed by a checker of a
third lineage (233 of 235 confirmed). The frozen analyser recovers \textbf{38 of
233} (16\%), with \textbf{no wrong answer}, 26 comparator-unresolved and 169
unanswered. Numeric facts largely survive (34 of 99; they are read by general
money and percentage rows), but names, sentences and yes/no facts do not (4 of 121):
the rule extractor produced 2.3 relation rows per template article and 0.6 per new
one, and the encoder, whose vocabulary has no relation type, cannot supply them.
The development figure measured the readers' fit to their templates; this one is the
figure that transfers.
\prov{out/experiments/inject-fresh-20260924/REPORT.md, RUN_NOTES.md, evaluation/summary.json}

A second, larger probe confirms it on new questions. For every firm we wrote four
free-form columns and executive interviews stating 20--26 answers the analyser gets
wrong or leaves unanswered (371 facts, 363 confirmed by an independent checker;
selection frozen before writing). The unchanged analyser recovers 24 of 363 (13
wrong, 12 unresolved, 314 unanswered). Eight of the thirteen errors were one
aggregate formula taking a team size (``a 14-person team'') for the firm's headcount;
restricting it to firm-wide totals removes them there and six certified errors on
the original record (reported above), with no certified correct answer lost.
Both probes place the gap in the same layer, reading natural prose, and it is a
learned extractor's to close, not further hand-written readers'.
\prov{out/experiments/expanded-20260925/REPORT.md; headcount-fix-20260925/README.md}

\paragraph{Question-anchored rows.}
The missing relation type is what the encoder's question-anchored rows supply
(\S\ref{sec:extraction}): a relation row and a metric row typed by the question
they answer, trained on labelled company pages and on free-form articles about
generated firms written by a third model, none from either probe, and read through
one generic clause that turns such a row into an answer. With them the analyser
recovers \textbf{64 of 233} (27\%) on the held-out articles, with 6 wrong, 32
unresolved and 131 unanswered, and \textbf{102 of 363} (28\%) on the second probe,
with 14 wrong and 13 unresolved, against 38 and 24 without them. No reader was
written for either probe. What it does not read is entities: of the 74
name facts of the second probe it answers 2, and of the 49 category facts 8,
where numbers and yes/no facts are read at 59 of 75 and 25 of 63. Four of the six errors were one defect in the number grammar
(a range separator read as a minus sign, a scale suffix not carried to the low end of
a range); with it fixed the probe gives 66 with 2 wrong, a development figure since
the defect was found on these errors. \S\ref{sec:comparison} reports these rows
on the full cohorts with the second probe's articles in the record.
\prov{out/experiments/v4-biz-encoder-eval-20260925/README.md (v4b\_fresh, v4b\_exp: clean; v4c\_fresh: after the range fix); v4-biz-encoder-16firms-20260925/DETAILS\_new\_cohort.md (by answer type)}

\subsection{Comparison with language-model readers}
\label{sec:comparison}
We ran four language models on the same sixteen cohorts. Each receives the firm's
entire public record in context, the same question list and the same instructions
(answer from the records only; \texttt{UNDETERMINED} when the evidence is
insufficient; cite record identifiers with exact quotations), and is scored by the
same comparator against the same key. Two run on a consumer machine (Gemma~4
26B-A4B; Llama~3.1 8B) and two are hosted references (Gemini~3.8 Flash; Mistral
Medium 3.5, available only through an API). Table~\ref{tab:comparison} scores every
reader by one script under one definition.

\begin{table*}[t]
\centering\small
\caption{Readers on the 1,842 scorable cells, all scored by one script with the
same comparator and key. ``Answered'' counts substantive answers only
(\texttt{UNDETERMINED} and a gate-closing \texttt{NOT\_APPLICABLE} are not answers).
Evidence-backed answers are, for our analyser, those with an L3 certificate, and for
a language model, those whose every cited quotation occurs verbatim in the cited
record; precision is correct over correct plus wrong, excluding comparator-unresolved
cases. ``Verbatim quote'' is the share of a model's cited answers whose every
quotation is verbatim. The lower block repeats the comparison on the same cohorts
with the 64 expansion articles added (\S\ref{sec:injection}); there our analyser
uses the encoder's question-anchored rows, and its answers read directly from a
relation or metric row are L2 evidence (91 right, 17 wrong), not counted as
evidence-backed.}
\label{tab:comparison}
\resizebox{\textwidth}{!}{%
\begin{tabular}{@{}lrrrrrll@{}}
\toprule
\rowcolor{cbzTint}
 & & & & \multicolumn{2}{c}{evidence-backed} & & \\
\cmidrule(lr){5-6}
\rowcolor{cbzTint}
reader & correct & answered & acc.\ answering & correct / wrong / unres. & precision & verbatim quote & runs locally\\
\midrule
majority-class guess & 479 & 998 & 48\% & --- & --- & --- & yes, CPU\\
\hl \textbf{our analyser} (evidence + labelled prior) & \textbf{512} & 1,018 & 50\% & 42 / 3 / 9 & \textbf{93\%} & by construction & \textbf{yes, CPU}\\
\hl \quad evidence only (no prior) & 50 & 82 & 61\% & 45 / 7 / 9 & 87\% & by construction & yes, CPU\\
Gemma 4 26B-A4B & 148 & 307 & 48\% & 90 / 93 / 50 & 49\% & 92\% & yes, GPU\\
Llama 3.1 8B & 185 & 345 & 54\% & 98 / 78 / 28 & 56\% & 70\% & yes, CPU/GPU\\
\midrule
\multicolumn{8}{@{}l}{\color{cbzModel}\emph{hosted references, not comparable: larger, closed, remote, and sent the whole record}}\\
Gemini 3.8 Flash (hosted) & 55 & 98 & 56\% & 54 / 20 / 23 & 73\% & 99\% & no\\
Mistral Medium 3.5 (hosted) & 212 & 508 & 42\% & 156 / 158 / 83 & 50\% & 79\% & no\\
\midrule
\multicolumn{8}{@{}l}{\emph{the same cohorts with the 64 expansion articles}}\\
majority-class guess & 479 & 998 & 48\% & --- & --- & --- & yes, CPU\\
\hl \textbf{our analyser} (evidence + labelled prior) & \textbf{610} & 1,044 & 58\% & 71 / 7 / 14 & \textbf{91\%} & by construction & \textbf{yes, CPU}\\
\hl \quad evidence only (no prior) & 164 & 235 & 70\% & 65 / 15 / 14 & 81\% & by construction & yes, CPU\\
Gemma 4 26B-A4B & 351 & 627 & 56\% & 257 / 145 / 108 & 64\% & 93\% & yes, GPU\\
Llama 3.1 8B & 440 & 827 & 53\% & 190 / 118 / 50 & 62\% & 52\% & yes, CPU/GPU\\
\cmidrule(l){1-8}
\multicolumn{8}{@{}l}{\color{cbzModel}\emph{hosted references, not comparable: larger, closed, remote, and sent the whole record}}\\
Gemini 3.8 Flash (hosted) & 254 & 406 & 63\% & 253 / 68 / 83 & 79\% & 100\% & no\\
Mistral Medium 3.5 (hosted) & 445 & 807 & 55\% & 356 / 186 / 124 & 66\% & 84\% & no\\
\bottomrule
\end{tabular}}
\prov{ai-libs/nn/cmcompare/comparison_table.py --capture out/experiments/headcount-fix-20260925/capture -> headcount-fix-20260925/comparison-table/table.json; lower block: --view injected on precision-fixes-20260926/capture_v4h_exp -> table_v4h_exp_injected/table.json and --llm-suffix _exp -> v4-biz-encoder-16firms-20260925/table_with_llms_injected/table.json}
\end{table*}

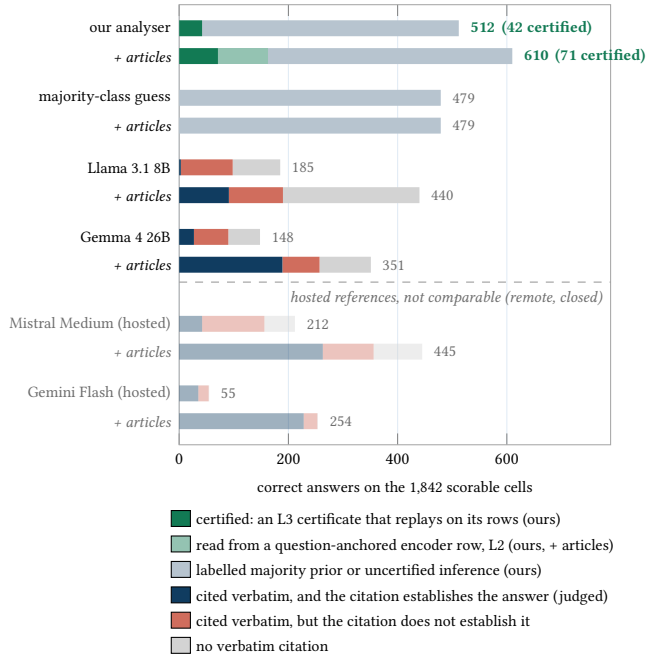
\begin{figure}[t]
\centering
\begin{tikzpicture}
\begin{axis}[xbar stacked, width=0.86\columnwidth, height=7.6cm, bar width=6pt,
  xmin=0, xmax=790, xtick={0,200,400,600}, clip=false, xlabel={correct answers on the 1,842 scorable cells},
  xlabel style={font=\scriptsize}, tick label style={font=\scriptsize},
  ytick pos=left, y tick style={draw=none}, ytick={12,11.2,10,9.2,8,7.2,6,5.2,3.5,2.7,1.5,0.7}, y=0.46cm, enlarge y limits=0.06,
  yticklabels={our analyser,{\itshape + articles},majority-class guess,{\itshape + articles},Llama 3.1 8B,{\itshape + articles},Gemma 4 26B,{\itshape + articles},\textcolor{cbzModel}{Mistral Medium (hosted)},{\color{cbzModel}\itshape + articles},\textcolor{cbzModel}{Gemini Flash (hosted)},{\color{cbzModel}\itshape + articles}},
  legend style={font=\scriptsize, at={(0.5,-0.13)}, anchor=north, draw=none, legend columns=1},
  legend cell align=left, axis line style={draw=cbzModel!60}, xmajorgrids, grid style={cbzTint}]
\addplot[fill=cbzCert, draw=none] coordinates {(42,12) (71,11.2) (0,10) (0,9.2) (0,8) (0,7.2) (0,6) (0,5.2) (0,3.5) (0,2.7) (0,1.5) (0,0.7)};
\addplot[fill=cbzCert, draw=none, fill opacity=0.4, forget plot] coordinates {(0,12) (0,11.2) (0,10) (0,9.2) (0,8) (0,7.2) (0,6) (0,5.2) (0,3.5) (0,2.7) (0,1.5) (0,0.7)};
\addplot[fill=cbzCert!45, draw=none] coordinates {(0,12) (91,11.2) (0,10) (0,9.2) (0,8) (0,7.2) (0,6) (0,5.2) (0,3.5) (0,2.7) (0,1.5) (0,0.7)};
\addplot[fill=cbzCert!45, draw=none, fill opacity=0.4, forget plot] coordinates {(0,12) (0,11.2) (0,10) (0,9.2) (0,8) (0,7.2) (0,6) (0,5.2) (0,3.5) (0,2.7) (0,1.5) (0,0.7)};
\addplot[fill=cbzInk!30, draw=none] coordinates {(470,12) (448,11.2) (479,10) (479,9.2) (0,8) (0,7.2) (0,6) (0,5.2) (0,3.5) (0,2.7) (0,1.5) (0,0.7)};
\addplot[fill=cbzInk!30, draw=none, fill opacity=0.4, forget plot] coordinates {(0,12) (0,11.2) (0,10) (0,9.2) (0,8) (0,7.2) (0,6) (0,5.2) (0,3.5) (0,2.7) (0,1.5) (0,0.7)};
\addplot[fill=cbzInk, draw=none] coordinates {(0,12) (0,11.2) (0,10) (0,9.2) (3,8) (91,7.2) (27,6) (189,5.2) (0,3.5) (0,2.7) (0,1.5) (0,0.7)};
\addplot[fill=cbzInk, draw=none, fill opacity=0.4, forget plot] coordinates {(0,12) (0,11.2) (0,10) (0,9.2) (0,8) (0,7.2) (0,6) (0,5.2) (42,3.5) (263,2.7) (35,1.5) (228,0.7)};
\addplot[fill=cbzFalse!80, draw=none] coordinates {(0,12) (0,11.2) (0,10) (0,9.2) (95,8) (99,7.2) (63,6) (68,5.2) (0,3.5) (0,2.7) (0,1.5) (0,0.7)};
\addplot[fill=cbzFalse!80, draw=none, fill opacity=0.4, forget plot] coordinates {(0,12) (0,11.2) (0,10) (0,9.2) (0,8) (0,7.2) (0,6) (0,5.2) (114,3.5) (93,2.7) (19,1.5) (25,0.7)};
\addplot[fill=cbzModel!30, draw=none] coordinates {(0,12) (0,11.2) (0,10) (0,9.2) (87,8) (250,7.2) (58,6) (94,5.2) (0,3.5) (0,2.7) (0,1.5) (0,0.7)};
\addplot[fill=cbzModel!30, draw=none, fill opacity=0.4, forget plot] coordinates {(0,12) (0,11.2) (0,10) (0,9.2) (0,8) (0,7.2) (0,6) (0,5.2) (56,3.5) (89,2.7) (1,1.5) (1,0.7)};
\draw[dashed, cbzModel!70] (axis cs:0,4.7) -- (axis cs:790,4.7);
\node[font=\scriptsize\itshape, text=cbzModel, anchor=north east] at (axis cs:790,4.7) {hosted references, not comparable (remote, closed)};
\node[font=\scriptsize\bfseries, text=cbzCert, anchor=west] at (axis cs:518,12) {512 \,(42 certified)};
\node[font=\scriptsize\bfseries, text=cbzCert, anchor=west] at (axis cs:616,11.2) {610 \,(71 certified)};
\node[font=\scriptsize, text=cbzModel, anchor=west] at (axis cs:485,10) {479};
\node[font=\scriptsize, text=cbzModel, anchor=west] at (axis cs:485,9.2) {479};
\node[font=\scriptsize, text=cbzModel, anchor=west] at (axis cs:191,8) {185};
\node[font=\scriptsize, text=cbzModel, anchor=west] at (axis cs:446,7.2) {440};
\node[font=\scriptsize, text=cbzModel, anchor=west] at (axis cs:154,6) {148};
\node[font=\scriptsize, text=cbzModel, anchor=west] at (axis cs:357,5.2) {351};
\node[font=\scriptsize, text=cbzModel, anchor=west] at (axis cs:218,3.5) {212};
\node[font=\scriptsize, text=cbzModel, anchor=west] at (axis cs:451,2.7) {445};
\node[font=\scriptsize, text=cbzModel, anchor=west] at (axis cs:61,1.5) {55};
\node[font=\scriptsize, text=cbzModel, anchor=west] at (axis cs:260,0.7) {254};
\legend{{certified: an L3 certificate that replays on its rows (ours)},{read from a question-anchored encoder row, L2 (ours, + articles)},{labelled majority prior or uncertified inference (ours)},{cited verbatim, and the citation establishes the answer (judged)},{cited verbatim, but the citation does not establish it},{no verbatim citation}}
\end{axis}
\end{tikzpicture}
\caption{Where each reader's correct answers come from. Ours are certified (L3,
replayable) or labelled prior; a language model's are split by a blind audit of
its verbatim citations (\S\ref{sec:comparison}): the citation establishes the answer,
or it is partial, unrelated or contradicting. The prior alone scores 479.
The second bar of each pair (\emph{+ articles}) is the same reader on the same cohorts
with the 64 expansion articles added (\S\ref{sec:injection}); ours there uses the
encoder's question-anchored rows, which answer questions directly; such an
answer is evidence (L2), not certified, because replaying it only repeats the encoder's reading.
Below the dashed rule, dimmed: hosted models, shown for reference only; they are larger,
closed and remote, and receive the whole record, so they are not the comparison.}
\Description{Stacked horizontal bars of correct answers per reader, two bars per reader (original cohorts and with articles), split by the source of the answer.}
\label{fig:sources}
\end{figure}

\paragraph{Reading the table.}
Among locally runnable readers our analyser confirms the most cells, but of its
512, 465 are the labelled majority prior (the guess alone scores 479) and 5 are
uncertified L1 inferences.
Counting a correct gate-closing \texttt{NOT\_APPLICABLE} as an answer inflates every
total (the guess to 988, our analyser to 1,011, Gemma~4 to 309, Mistral to 512)
without changing the ordering of evidence-backed precision; we report the stricter
definition. The fair comparison is on the answers each reader backs with evidence.
Our certified answers are few (54 of 2,048 slots, 2.6\%) and right in 42 of 45
resolved cases (93\%); above the display threshold of 0.5, 45 remain, 37 correct,
none wrong and 8 unresolved. The three errors, all below the threshold, come from
one aggregate formula (the most frequent posting family taken as the function with
the most open roles), whose proof replays correctly on its rows: the rows were read
correctly and the formula answers a different question, a defect the certificate
localises to one clause. A second formula of this kind, the latest headcount
statement taken as the firm's total, produced six further errors until it was
restricted to statements of a firm-wide total (the others were team sizes). No
certified reader or rule answer is wrong. Every L3 certificate replays, and an
independent audit of the 49 comparable certified cells finds none insufficient (40
supported, 4 redundant, 5 prior-coincident). The language models make many more
quote-backed answers (97--397) at lower precision (49--73\%), and 1--30\% of their
cited answers carry a quotation that is not verbatim.
\prov{out/experiments/provenance-20260924/REPORT.md, ANSWERS.csv; support-check-certs-20260924/summary.json; evidence-audit-20260925/audit.json, AUDIT.csv}

\paragraph{Forward derivation versus post-hoc attribution.}
The difference is one of mechanism, not only of rate. In our analyser, evidence is a
\emph{forward} input: an answer is computed from recorded rows, and its certificate
replays the clause on exactly those rows, so a citation cannot be decorative by
construction. A language model produces the answer and its citation together;
nothing in the mechanism makes the answer follow from the quotation, so whether the
cited text supports the answer is incidental, and the quotations are mostly verbatim
while the \emph{link} is not. On the 359 cells of \S\ref{sec:unsupported} where the
base-rate answer to a yes/no question is ``no'' and correct, Mistral Medium gives 40
wrong answers with verbatim quotations against 32 right ones, and Gemma~4 gives 26
against 5. A blind audit of every cited, substantive answer of the four models
(1,131 cells; each judged with the question, the answer and the full cited records,
without the key) makes the gap measurable: of the correct answers whose citations
are verbatim, the citation \emph{establishes} the answer in 35 of 54 for Gemini
(65\%), 42 of 156 for Mistral (27\%), 27 of 90 for Gemma~4 (30\%) and 3 of 98 for
Llama (3\%); the rest are partial (on topic, not sufficient), unrelated (a sentence
on last year's accounts cited for ``no funding round is being prepared'', or an
argument from silence) or contradicting; a second judge agrees with the first on
the supports/does-not split for 109 of a random 120 cells ($\kappa=0.78$).
Figure~\ref{fig:sources} shows the split.
Such answers are the base-rate guess presented as evidence; our analyser labels
the same guess as a guess, and each of its 42 certified correct answers rests on the
rows its proof replays. Forward derivation guarantees the link, not the truth: our
three certified errors are correct links through a wrong formula. The measured counterpart of our sufficiency replay,
for a language model, is to re-ask each question with only the quotation the model
cited and count the answers that no longer follow; we describe that measurement in
\S\ref{sec:limits} as the next step.
\prov{out/experiments/premise-gap-20260924/llm_on_cells.py}

\paragraph{On facts the record supports.}
The injection protocol gives cells whose support is known. Asking every reader the
same twenty questions over the injected cohorts, of 235 supported facts our analyser
recovers 188 (80\%), Mistral Medium 171 (73\%), Llama~3.1 145 (62\%), Gemini~3.8
137 (58\%) and Gemma~4 120 (51\%); the models leave 46--64 answers unresolved by the
comparator, against our 33. This is a \emph{development} result for our analyser:
its dedicated readers were specified from this probe's own diagnostics, whereas the
models answered zero-shot; the held-out version of the probe, with the analyser
frozen and the articles in unseen phrasing, is reported in \S\ref{sec:injection}.
\prov{decoy188 transfer_raw/*__undefended__*.json; inject20-readersv2 summary; inject-fresh-20260924/PROTOCOL.md}

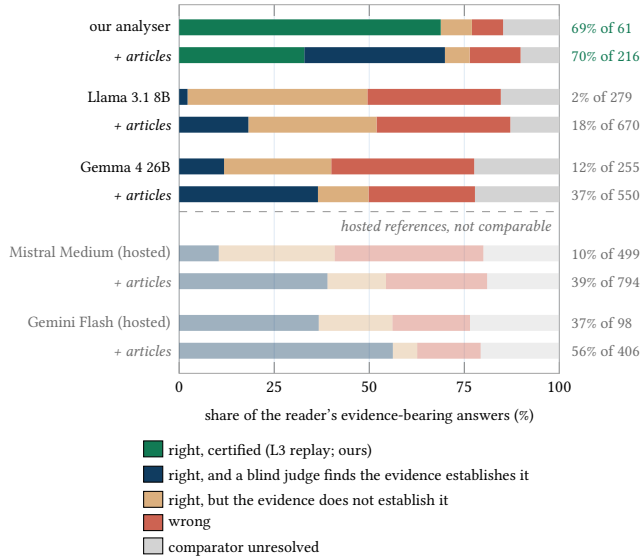
\begin{figure}[t]
\centering
\begin{tikzpicture}
\begin{axis}[xbar stacked, width=0.78\columnwidth, height=6.6cm, bar width=6pt,
  xmin=0, xmax=100, xtick={0,25,50,75,100}, clip=false,
  xlabel={share of the reader's evidence-bearing answers (\%)},
  xlabel style={font=\scriptsize}, tick label style={font=\scriptsize},
  ytick pos=left, y tick style={draw=none}, ytick={10,9.2,8,7.2,6,5.2,3.5,2.7,1.5,0.7}, y=0.46cm, enlarge y limits=0.07,
  yticklabels={our analyser,{\itshape + articles},Llama 3.1 8B,{\itshape + articles},Gemma 4 26B,{\itshape + articles},\textcolor{cbzModel}{Mistral Medium (hosted)},{\color{cbzModel}\itshape + articles},\textcolor{cbzModel}{Gemini Flash (hosted)},{\color{cbzModel}\itshape + articles}},
  legend style={font=\scriptsize, at={(0.42,-0.16)}, anchor=north, draw=none, legend columns=1},
  legend cell align=left, axis line style={draw=cbzModel!60}, xmajorgrids, grid style={cbzTint}]
\addplot[fill=cbzCert, draw=none] coordinates {(68.8,10) (32.9,9.2) (0,8) (0,7.2) (0,6) (0,5.2) (0,3.5) (0,2.7) (0,1.5) (0,0.7)};
\addplot[fill=cbzCert, draw=none, fill opacity=0.4, forget plot] coordinates {(0,10) (0,9.2) (0,8) (0,7.2) (0,6) (0,5.2) (0,3.5) (0,2.7) (0,1.5) (0,0.7)};
\addplot[fill=cbzInk, draw=none] coordinates {(0,10) (37,9.2) (2.2,8) (18.2,7.2) (11.8,6) (36.5,5.2) (0,3.5) (0,2.7) (0,1.5) (0,0.7)};
\addplot[fill=cbzInk, draw=none, fill opacity=0.4, forget plot] coordinates {(0,10) (0,9.2) (0,8) (0,7.2) (0,6) (0,5.2) (10.4,3.5) (39,2.7) (36.7,1.5) (56.2,0.7)};
\addplot[fill=cbzGuess!55, draw=none] coordinates {(8.2,10) (6.5,9.2) (47.3,8) (33.7,7.2) (28.2,6) (13.3,5.2) (0,3.5) (0,2.7) (0,1.5) (0,0.7)};
\addplot[fill=cbzGuess!55, draw=none, fill opacity=0.4, forget plot] coordinates {(0,10) (0,9.2) (0,8) (0,7.2) (0,6) (0,5.2) (30.5,3.5) (15.4,2.7) (19.4,1.5) (6.4,0.7)};
\addplot[fill=cbzFalse!85, draw=none] coordinates {(8.2,10) (13.4,9.2) (35.1,8) (35.2,7.2) (37.6,6) (28,5.2) (0,3.5) (0,2.7) (0,1.5) (0,0.7)};
\addplot[fill=cbzFalse!85, draw=none, fill opacity=0.4, forget plot] coordinates {(0,10) (0,9.2) (0,8) (0,7.2) (0,6) (0,5.2) (39.1,3.5) (26.6,2.7) (20.4,1.5) (16.7,0.7)};
\addplot[fill=cbzModel!30, draw=none] coordinates {(14.8,10) (10.2,9.2) (15.4,8) (12.9,7.2) (22.4,6) (22.2,5.2) (0,3.5) (0,2.7) (0,1.5) (0,0.7)};
\addplot[fill=cbzModel!30, draw=none, fill opacity=0.4, forget plot] coordinates {(0,10) (0,9.2) (0,8) (0,7.2) (0,6) (0,5.2) (20,3.5) (19,2.7) (23.5,1.5) (20.7,0.7)};
\draw[dashed, cbzModel!70] (axis cs:0,4.7) -- (axis cs:100,4.7);
\node[font=\scriptsize\itshape, text=cbzModel, anchor=north east] at (axis cs:100,4.7) {hosted references, not comparable};
\node[font=\scriptsize, text=cbzCert, anchor=west] at (axis cs:101,10) {69\% of 61};
\node[font=\scriptsize, text=cbzCert, anchor=west] at (axis cs:101,9.2) {70\% of 216};
\node[font=\scriptsize, text=cbzModel, anchor=west] at (axis cs:101,8) {2\% of 279};
\node[font=\scriptsize, text=cbzModel, anchor=west] at (axis cs:101,7.2) {18\% of 670};
\node[font=\scriptsize, text=cbzModel, anchor=west] at (axis cs:101,6) {12\% of 255};
\node[font=\scriptsize, text=cbzModel, anchor=west] at (axis cs:101,5.2) {37\% of 550};
\node[font=\scriptsize, text=cbzModel, anchor=west] at (axis cs:101,3.5) {10\% of 499};
\node[font=\scriptsize, text=cbzModel, anchor=west] at (axis cs:101,2.7) {39\% of 794};
\node[font=\scriptsize, text=cbzModel, anchor=west] at (axis cs:101,1.5) {37\% of 98};
\node[font=\scriptsize, text=cbzModel, anchor=west] at (axis cs:101,0.7) {56\% of 406};
\legend{{right, certified (L3 replay; ours)},{right, and a blind judge finds the evidence establishes it},{right, but the evidence does not establish it},{wrong},{comparator unresolved}}
\end{axis}
\end{tikzpicture}
\caption{How much of each reader's evidence is correct evidence. Each bar is 100\% of the
reader's evidence-bearing answers (ours: every answer that cites evidence, L1 to L3; a
language model: every substantive answer with a citation); the label gives the share that is right
\emph{and} rests on evidence that establishes it, and the number of such answers.
Language-model citations and our uncertified (L1, L2) answers were judged blind with one
rubric (\S\ref{sec:comparison}). Second bar of each pair: the cohorts with the 64
expansion articles. Below the dashed rule, dimmed: hosted models, for reference only.}
\Description{Stacked horizontal percentage bars per reader showing the share of evidence-bearing answers that are right on establishing evidence, right without it, wrong, or unresolved.}
\label{fig:evshare}
\end{figure}

\paragraph{With the expansion articles.}
The lower block of Table~\ref{tab:comparison} repeats the comparison on the same
cohorts with the 64 expansion articles of \S\ref{sec:injection} (plain-prose columns
and interviews stating 363 facts); every model was re-run on the new records and its
citations re-audited blind (2,420 cited answers; a second judge agrees on the
supports/does-not split for 95\% of a random 240, $\kappa=0.89$). The picture
changes. Every model gains 199--255 correct answers, and its evidence becomes real:
the citation establishes the correct answer in 228 of 253 verbatim-cited correct
answers for Gemini (90\%), 263 of 356 for Mistral (74\%), 189 of 257 for Gemma~4
(74\%) and 91 of 190 for Llama (48\%). Our analyser gains 98 (512 to 610, prior
guesses included) with the question-anchored rows, which answer questions
directly; such an answer is L2 evidence, not certified, because replaying
it only repeats the encoder's reading. The gains sit where the articles are. On the
363 facts they state, our evidence-bearing answers (L3 and L2) are 106 right and 8
wrong, against answers with an establishing citation of 254 right and 22 wrong for
Mistral, 188 and 22 for Gemini, 162 and 29 for Gemma~4 and 112 and 2 for Llama. On
the rest of the record our certified answers remain the most precise evidence (47
right, 3 wrong; our encoder-row answers there are 9 right and 13 wrong), where the models' establishing citations are 56 right and 55 wrong
(Mistral), 40 and 26 (Gemini), 39 and 30 (Gemma~4) and 10 and 3 (Llama), many of
the wrong ones taken from incidental claims in the articles that contradict the key.
On plain prose a general model reads more than our analyser; what it lacks is a
guarantee that its citation is the reason for its answer, which our certificate
gives where it applies. Our figures follow four fixes to the encoder's rows (a topic
and unit guard on metric rows, readable numbers, L2 for encoder-row answers, symbolic
readings ranked first) that were developed on these articles, so they are a
development result; the encoder was not trained on them.
Figure~\ref{fig:evshare} puts every reader on one scale: of all its
\emph{evidence-bearing} answers (for ours, every answer that cites evidence, L1 to
L3; for a model, every substantive answer with a citation), the share that is right
\emph{and} rests on evidence that establishes it. Our uncertified answers (L1, L2)
were judged blind with the models' rubric, all 131 of them twice: the encoder row
establishes 80 of the 91 right L2 answers for the first judge and 82 for the second,
who agree on the supports/does-not split for 128 of 131 ($\kappa=0.95$). With the articles, 70\% of our
216 evidence-bearing answers are right on establishing evidence (71 certified, 80
judged), against 56\% of Gemini's 406, 39\% of Mistral's 794, 37\% of Gemma's 550 and
18\% of Llama's 670; on the original cohorts the shares are 69\% of our 61 against
2--37\% for the models.
\prov{out/experiments/precision-fixes-20260926/README.md; v4-biz-encoder-16firms-20260925/EVIDENCE_ASSESSMENT.md; evidence-audit-exp-20260926/, evidence-audit-exp2-20260926/ (second_judge/AGREEMENT.md); evidence-audit-ours-20260926/audit.json, evidence_share.json}

\paragraph{Cost and locality.}
One pass over the sixteen firms cost \$0.14 (Gemma), \$0.03 (Llama), \$2.73
(Gemini) and \$2.32 (Mistral) through a hosted provider (with the articles: \$0.13,
\$0.03, \$3.05 and \$2.63), and sends every firm's public record to a third party. Our analyser runs on the defender's machine at no
marginal cost. For a defensive tool handling a client's material, locality is a
requirement rather than a convenience.

\subsection{A real company}
\label{sec:realfirm}
We ran the analyser on Apple's public footprint as collected by our crawler: 44
documents and 1,883 passages (registry profile, company pages, press and
encyclopaedic text). No key exists for a real company, so we report coverage and
inspect answers; accuracy is not claimed. Registry field mappings alone answer 3 of
the 128 questions; rules and solvers answer 6; rules, our encoder and solvers answer
10, every one of them with an L3 certificate; the statistical classifier and the
majority prior refuse to answer, because they are fitted only on synthetic firms,
which is the guard working as intended. The encoder adds 3,475 evidence rows to the
rules' 2,298 and lifts coverage from 6 to 10 questions, including the stated
objective, the primary technology stack and a critical supplier dependency, two of
them from the additive keys.
\prov{out/experiments/provenance-offsets-20260924/apple/}

Two findings temper the number. First, the dedicated readers that recover 188 of 235
injected facts answer almost nothing here: the rule extractor produces none of the
relation rows they consume on Apple's 1,883 passages, where it produced them readily
on regular synthetic prose. The question-anchored rows of \S\ref{sec:injection}
were trained to supply them, and on this crawl they do not: with them coverage rises
from 10 to 13 questions, but the encoder produces 2 relation rows in 1,883 passages,
and of the four answers read directly from its 66 metric rows three are wrong on
inspection (an index-fund ownership share taken as the largest owner's, a labour-cost
share as the largest customer's, the Apple~II's installed base as today's paying
customers) and one is at best partial; the two new certified answers (revenue growth,
a store opening) come from the symbolic readers, and two of the ten answers without
these rows are lost, because the heads trained with them emit half as many rows of
the older types; keeping both row sets loses none of the ten and adds exactly the
three wrong answers. The entity gap of \S\ref{sec:injection} is a real-text gap too. Second, two of the ten answers are plainly wrong, the target named as its own
primary competitor and a word from a contract sentence returned as its largest
customer, both from the graph solver's entity choice. Their certificates replay,
which illustrates what a certificate does and does not establish: the rows were
read and the clause reproduces the answer, but a type check on arguments (a
competitor must be another organisation; a customer must be an entity) is needed to
reject both.

\paragraph{Infrastructure exhaust.}
Two public channels name a firm's suppliers and systems without a sentence being
written: DNS records and certificate-transparency logs. A domain-verification token
in a TXT record is published at a vendor's request and names that vendor
\cite{vandertoorn2020txting}; the SPF policy and MX records name who sends and
receives the firm's mail \cite{liu2021mail}; and every hostname a certificate was
issued for is logged, so product, environment and tenant tokens in hostnames, and
their issue dates, are public \cite{roberts2019transparency}. The core reads these
as declared fields (\S\ref{sec:extraction}): a token becomes a \emph{vendor} row
naming the vendor, never the token value; a hostname becomes a \emph{host-hint} row
with its environment, role, region and name labels. On the same two crawls, DNS
alone names 13 vendors for Apple (Adobe, Atlassian, Cisco Webex, Dynatrace, Oracle
Cerner, \ldots; mail self-hosted) and 23 for Palantir (Anthropic, OpenAI, Cursor,
Stripe, Shopify, HackerOne, OneTrust, \ldots; mail through Proofpoint), none stated
in either firm's text. Palantir's certificate log enumerates 483 numbered
\texttt{foundry-\textit{region}-\textit{n}.status} hosts across 19 region codes,
whose first-seen curve (by validity start, from a 500-certificate sample) runs from
38 in October 2023 to 483 in August 2026 with steps of +143 and +76: a deployment
count and its growth, published by the status page's certificates. Ten further
hostnames carry a name token (\texttt{velvet}, \texttt{regalia},
\texttt{concierge-staging}, \ldots) that occurs in none of the 9,476 passages; the
dossier reports them as \emph{leads} for the unannounced-initiative question with
their certificate, never as answers. The same run diffs successive snapshots of one
page into removed and added statements, following the observation that dropping a
metric or rewording a policy is itself disclosure
\cite{cohen2020lazy,biddle2024openai}; the one page crawled twice was unchanged,
so this reader has a unit test but no real positive yet. The frozen benchmark is
untouched by these keys: all 64 prediction files replay with identical answers.
\prov{out/experiments/infra-20260925/README.md}

\subsection{Defence: constrained interventions and transfer}
\label{sec:defence}
The analyser's citations make the defender's first question exact: \emph{which
public items carry this fact}. Removing a minimal sufficient set of carriers is not
enough when several independent sets exist; the actionable object is a minimum
hitting set of all sufficient sets~\cite{reiter1987diagnosis}. The defender's
problem is then a constrained synthesis in the sense of the privacy
funnel~\cite{makhdoumi2014funnel}: minimise the adversaries' recovery subject to
utility and the ledger of \S\ref{sec:problem}. The loop of measuring what an
adversary can read, changing the record and measuring again has a precedent in
operations security: an audit of routine flight filings that had let an adversary
anticipate a third of bombing missions, followed by changes that cut the share to one
in twenty~\cite{nsa1993purpledragon}.

Evidence from our earlier work on personal footprints shapes the design: adding
contradictory material does not overturn a well-grounded inference, whereas
displacing its carriers does. We therefore test three deterministic interventions,
none generated by a language model. \emph{Generalise} (admissible): the carrier
states a true but coarser value, a role instead of a name, a wider band that
contains the truth, a coarser window. \emph{Suppress} (conditional): the carrier
sentence is removed. \emph{Append-only control} (needs sign-off): carriers are left
untouched and neutral true sentences are added, which we expect to hide nothing. A
planner hides all of a firm's facts together, prefers generalisation, and escalates
a fact to suppression only when the analyser still recovers it.

\paragraph{Experiment.}
The secrets are the 188 injected facts the analyser recovers
(\S\ref{sec:injection}), whose carriers are known. We measure the hide rate and cost
(characters changed) against our analyser, collateral changes to other answers, and
\emph{transfer}: the share of secrets each language-model adversary recovered on the
undefended cohorts that it no longer recovers after the defence, together with any
secret it recovers only after the defence. Table~\ref{tab:defence} gives the result.

\begin{table*}[t]
\centering\small
\caption{Defence on the 188 secrets. Columns 2 and 3: our analyser; columns 4--7: the
secrets each adversary still recovers, with the transfer hide rate (the share of its
own undefended recoveries it no longer recovers).}
\label{tab:defence}
\resizebox{\textwidth}{!}{%
\begin{tabular}{@{}lrrrrrr@{}}
\toprule
\rowcolor{cbzTint}
condition (ledger class) & ours & chars changed & Gemma 4 & Llama 3.1 & Gemini 3.8 & Mistral 3.5\\
\midrule
undefended & 188 & --- & 114 & 134 & 126 & 158\\
\hl generalise + 2 escalations (admissible) & \textbf{0} & \textbf{12,612} & 0 (100\%) & 1 (100\%) & 0 (100\%) & 0 (100\%)\\
suppress (conditional) & 0 & 21,111 & 1 (100\%) & 0 (100\%) & 0 (100\%) & 0 (100\%)\\
append only (sign-off) & 188 & 10,252 & 119 & 147 & 134 & 162\\
\bottomrule
\end{tabular}}
\prov{research/counterbiz/decoy/experiments/decoy188-20260923T213829Z/result.json}
\end{table*}

Generalisation alone hid 186 of 188 secrets from our analyser; the planner escalated
two to suppression, and both turned out to be a rule-extraction defect (``whether a
viable substitute exists'' read as an affirmation) rather than a leak in the
rewrite. Generalisation and suppression hid every secret from all four adversaries,
but generalisation edited 40\% fewer characters and leaves a true, coarser statement
public instead of a conspicuous gap, which matters because a gap can itself be
read~\cite{grossman1981,milgrom1981}. Appending neutral true sentences hid nothing
from our analyser; the adversaries' apparent 4--20\% hide rates under append were
offset by 10--33 secrets they newly recovered in the same condition, run-to-run
variation, replicating on organisations the finding that additive material does not
displace a grounded inference. Collateral changes to non-secret answers were 3
(generalise) and 8 (suppress). One partial leak survived: a context sentence kept
true and unchanged let the adversaries approximate a hidden hiring capability, so a
generalisation rule must cover the secret's context, not only its carrier. The
experiment cost \$4.25 in adversary queries.

\paragraph{What this does not show.}
Each secret here has exactly one carrier and no other support, by construction, so
editing that carrier necessarily hides it. Full transfer is the expected outcome of
an easy case, not evidence that planning against a transparent analyser transfers in
general. The realistic case, a fact supported by several independent records where
the planner must find a minimum hitting set and the adversary may exploit residual
context, is the next experiment, and the redundant-support cells that the
certificates already identify (\S\ref{sec:certificates}) are its natural secrets. A
second caution comes from a documented case: when investors estimated a retailer's
sales from its sequential order numbers, the firm began skipping numbers, and the
skipping was itself detected by timed counts~\cite{luckin2020report}. Altering a
leaky signal is a signal, so a defence must be measured against an adversary who
knows the record has changed.

\section{Case study: individuals}
\label{sec:personal}

Our work began on individuals: what a person's own data exports and public activity
reveal to an adversary, and which changes to that footprint reduce it. The personal
pack reuses the core of \S\ref{sec:architecture} unchanged; the organisation
analyser's 57,344 saved answers and their certificates replay byte-identically after
the pack was added. This section reports development measurements on synthetic
people. The rules were designed while inspecting the same cohorts, so none of the
numbers below is a held-out result.

\subsection{What differs from organisations}
An organisation's private facts are often \emph{stated} in its footprint, a
headcount in a filing or a supplier in a case study, which is why extraction and
dedicated readers carry the organisation case. A person's sensitive attributes are
rarely stated; they are \emph{patterns} across many small items: a routine across
timestamps, a residence across location traces, a leaning across likes and searches.
Four consequences shape the pack. Statistical and aggregate solvers over activity
logs carry many answers, so they must record their participating rows to reach L3.
Personal text mentions family, friends and public figures as often as the author and
is less literal (quotation, reposts, humour), so every row carries a subject (owner,
named other, public figure) and a register field, and no answer is ever emitted
about a third party. A personal fact typically has many carriers, so the defence
works on hitting sets from the start, with the actions platforms allow an owner
(delete, redact, restrict visibility) rather than rewording a document. And no real
personal data is used: the benchmark is synthetic, and real data may enter only
through a consented study in which each volunteer's system is erased after the
session.

\subsection{Instrument and cohorts}
Sixteen synthetic personas, one per adversary archetype, each with a complete
private specification over 44 dimensions in seven modules (identity, device and
operational security, rhythm, vulnerability, relationships, space, concealment and
history). The frozen question bank has 98 questions, 97 of them active, giving 1,552
slots per cohort, of which the frozen key (select-then-derive construction)
settles 1,515. The same sixteen people are rendered by different generators into
five cohorts in two access conditions, which we keep separate: three \emph{public}
footprints (posts, directories, news; one each from three generator lineages) and
two \emph{private-export} footprints (a stolen or compelled data archive with
searches, messages, calendar and location history, thirty days of activity). The
five cohorts are versions of the same sixteen people, not eighty independent
people. The comparator gives numeric tolerances from each question's definition and
declared option aliases, reports ordinal partial credit separately, and leaves
ambiguous free text unresolved.

\subsection{Where the score comes from}
The majority-class reference, fitted for each person on the other fifteen only,
answers \textbf{798 of the 1,515 settled cells fully correctly (52.7\%)}, with 101
partially correct and 111 unresolved. The organisation reference answers 479 of
1,842 (26.0\%). The instruments and populations differ, so this is not a matched
comparison, but the direction is the one \S\ref{sec:measurement} anticipates: base
rates are far more predictive of people than of firms, and the guessing term of
Equation~\ref{eq:split} is correspondingly larger. A reader that is scored against
private truth on this bank earns half its score before reading anything.

Language-model readers from earlier work on the same cohorts, pooled per cohort,
recover between 806 and 1,054 cells in the cohorts where they cover all sixteen
personas (Table~\ref{tab:personal}). This is an oracle union over several readers,
not a single reader's score, and it does not establish that a correct answer came
from the evidence. Compared on the same cells, the pooled readers beat the reference
by 8 to 256 cells; one question (the person's five closest contacts) is never
recovered in any cohort.

\begin{table}[t]
\centering\small
\caption{The personal instantiation on five cohorts of the same sixteen synthetic
people (1,515 settled cells each). Prior: leave-one-persona-out majority reference.
Pooled: union of earlier language-model readers (covering 4 of 16 personas for the
first cohort). Evidence: correct L3 answers from our readers and aggregates, before
$\rightarrow$ after the evidence-reading extension, from the rule readers and
statistical solvers alone; the personal head set of the encoder (below) is evaluated
against teacher labels only and is not part of these numbers. Selected: what the
complete pipeline returns, including labelled guesses.}
\label{tab:personal}
\resizebox{\columnwidth}{!}{%
\begin{tabular}{@{}llrrrr@{}}
\toprule
cohort & access & prior & pooled & evidence & selected\\
\midrule
generator A & public & 798 & 248$^{\ast}$ & 19 $\rightarrow$ 49 & 798 $\rightarrow$ 804\\
generator B & public & 798 & 806 & 15 $\rightarrow$ 30 & 798 $\rightarrow$ 798\\
generator C & public & 798 & 1,054 & 11 $\rightarrow$ 43 & 798 $\rightarrow$ 805\\
generator D & private export & 798 & 1,007 & 35 $\rightarrow$ 35 & 798 $\rightarrow$ 798\\
generator E & private export & 798 & 953 & 35 $\rightarrow$ 35 & 798 $\rightarrow$ 798\\
\bottomrule
\end{tabular}}
\par\smallskip{\footnotesize $^{\ast}$ readers cover 4 of 16 personas.}
\prov{research/counterpriv/out/experiments/personal-measurement-20260924-both-final/REPORT.md; personal-p5-20260924/REPORT.md; organisation reference 479: countermodel/out/experiments/comparison-table-20260924/table.json}
\end{table}

\subsection{Extraction, solvers and certificates}
Deterministic rules turn the 22,300 input records of the five cohorts into 52,660
evidence rows: 20,189 exact text spans and 32,471 native fields (timestamps,
coordinates, device and account fields), in 21 observation keys that map every one
of the 44 dimensions to the evidence that could support it. An extension adds
owner-profile fields, profile-name observations, identity links to directory
listings, and clause-level rows that keep quotations and reported speech apart from
the owner's own statements, for 73,286 rows in all; most additions are clause rows,
not new facts. No model is called at any point.

Readers cover 32 questions and aggregate or statistical solvers ten, three of them
also covered by a reader; together 39 of the 97 questions have an evidence solver. Readers answer explicit statements
(age, residence, profession, partner status, children, education, email provider),
with parent-based gates for not-applicable answers. The statistical solvers compute
platform rankings, daily first and last activity, weekday--weekend divergence,
repeated office presence, night-location clusters and mobility radius, abstaining on
sparse logs. An identity link is admitted only when a name from an owner profile
matches a directory record, and the proof then consumes both rows, so that removing
the identity evidence removes the answer. Routing admits a solver for a question only
if it beats the reference on the other people, with every version of a person held
out together. The independent verification replays every L3 proof: none is
insufficient, and no answer targets a third party.

\paragraph{Results.}
On public footprints, evidence reading works. Correct evidence answers rise from
19 to 49, 15 to 30 and 11 to 43, mostly residence country and city, profession and
completed education, and the selected score rises by 6 and 7 in two cohorts with no
previously correct answer lost. The new reader is right on 121 of its 131 public
attempts, wrong on 6 and unresolved on 4. Its two regressions show what remains: in
both, ``my wife'' sits in a sentence that denies something else, and the clause-level
reader suppresses the whole clause, which calls for predicate-level negation scope.
On private exports, the explicit readers add nothing: their few attempts are wrong,
and the 35 correct evidence answers per cohort come from the statistical solvers.
This is the harder and more realistic condition, a person's archive rather than
their public posts, and it is where a learned extractor adapted to informal text is
needed next.

\paragraph{A personal head set on the same encoder.}
The shared encoder was given a second head set over the personal pack's sixteen prose
keys, with subject (self, named other, public figure, unknown), the pack's assertion
statuses and the in-text register; third parties are recorded by name and relation
only, never profiled. Training text is synthetic and comes from 78 personas outside
the frozen sixteen; passages that a consumer-subscription model had written were
replaced by text from a model whose terms allow it, and non-Latin passages were
removed (4,160 passages, plus 1,432 denser passages that state concrete personal
facts). Labels come from an open-weight teacher (GLM-5.2, MIT), chosen by a bake-off
against a stronger reference (row F1 0.67 against 0.63 and 0.53 for two alternatives;
assertion agreement 0.83 against 0.61). Fine-tuning ran on a laptop CPU in about two
hours, because personal passages are short (49 tokens on average against 616 for
company pages). On the 1,500 public-tier passages of the frozen sixteen, against
teacher labels, the encoder reaches row F1 0.52 (precision 0.49, recall 0.55), the
rules 0.09 (recall 0.05); subject agreement is 0.99 against 0.07 for the rules,
which rarely attribute a mention to anyone. Occupation, the most frequent sensitive
key, goes from 0.00 to 0.70 and relationship mentions from 0.12 to 0.40; health,
belief and concealment remain thin (38, 1 and 8 reference rows) and are not yet
reliable. Labels from a weaker teacher on the less dense pool gave 0.29, so both the
teacher and the density of the training text mattered.
\prov{research/counterpriv/out/experiments/personal-encoder-eval-20260925/v2_full_glm52/EVAL.md; personal-teacher-bakeoff-20260925/README.md; personal-encoder-20260925/PASSAGES_V2.md, SPEND_20260925.md}

\subsection{Defence on personal footprints}
Two measurements test whether changing the footprint removes an inference. In the
first, each person receives two independent truthful records stating the number of
their children and their email provider; the analyser recovers all 32 target cells
per cohort. Deleting one carrier leaves all 32 recoverable, because the second
carrier still supports each answer; deleting or redacting the minimum hitting set of
discovered supports leaves no reader answer. Restricting visibility removes the
readings from public views but not from a full private export, and an adversary's
previously obtained copy is untouched by any action. The prior still guesses 22 of
the 32 correctly, so removing a source path does not remove a guess.

The second measurement uses routines already present in the private logs (typical
wake and sleep hours) and a second, independent estimator held in reserve. Planned
deletions of about 2.6 records per person, 1.1\% of accepted records, with messages,
email and calendar protected, lower the primary estimator's correct answers from 21
to 13 of 32, while random deletions of equal cost change nothing. The reserve
estimator falls only from 16 to 15. The plans disrupt the analyser they were planned
against and barely transfer to another. On organisations, rewriting single carriers
transferred fully to four language models (\S\ref{sec:defence}); for statistical
personal inferences, planning against one estimator does not protect against a
second, and general protection is not established.
\prov{research/counterpriv/P3.md, P4.md; out/experiments/personal-p3-20260924, personal-p4-20260924}

\subsection{Decoy generation without a language model}
\label{sec:decoy}
Deletion is not always available: an owner cannot delete records already exported, and
many platforms keep activity logs. The alternative is to \emph{add} activity that moves
an inference. We separate decoy generation into two layers. A \emph{planner} decides
the ingredients of the signal: which inference to move, which kind of record to add,
at which times, on how many days and within what budget. A \emph{renderer} turns those
ingredients into content a platform would accept, such as the text of a post or the
wording of a search. The planner is where the privacy effect is decided and measured;
the renderer is where a generative model is useful in a product, because inventing
plausible, varied content is exactly what such models do well, and there it is gated
by the planner's specification rather than trusted to choose the signal
(\S\ref{sec:architecture}). This section tests the planner alone, with deliberately
plain rendering, and asks one question: can a deterministic planner, with no language
model, generate additions that weaken an inferred signal?

\paragraph{Planner.}
The target is the pair of routine questions of the previous experiment, typical wake
and sleep hour, which the primary estimator reads as the modal hour of each day's first
and last activity. Given the observable export and the estimator's current answers,
never the key, the planner searches a declared grid: early hours (04:00, 05:00) and late
hours (01:00--03:00), and 7, 14, 21 or 30 days or all observed days, within a budget of
20\% of the person's original records and at most 60 additions. It chooses the schedule
that moves the most current estimates outside the question's tolerance, preferring a
concrete alternative answer to an abstention and then fewer additions. Each addition is
a search event from a small set of neutral, non-personal queries. An append-only
transaction checks the source hash, owner scope, record schema, timestamps, duplicates
and budget, and keeps every original record unchanged. Plans were frozen before the
second estimator was run, and predictions before either answer key was loaded.

\paragraph{Results.}
The planner adds 226 records per private cohort, 14.1 per person. The primary
estimator's correct answers fall from 21 to 11 of 32: thirteen correct answers are
lost and three wrong ones become correct. The same number of additions at random times
changes almost nothing (21 to 20), so the effect comes from the plan, not the volume.
For one persona, the estimates move from 07:00 and 23:00 to 04:00 and 01:00 with 27
additions. So the answer to the question is yes: a deterministic planner weakens the
signal it targets, cheaply and without touching the genuine record.

It does not transfer. The median estimator, which the planner never queried, rises from
16 to 18 correct: six correct answers are lost and eight are gained, because additions
that move a mode can pull a median towards the truth. An observer holding an earlier
copy, or one that can identify every added record, recovers the original answers
exactly. The full routed system is unchanged at 798, because for these two questions it
still answers with the population prior, which is right on 18 of 32 regardless of the
log. Planning against one estimator therefore gives a planner that works against that
estimator; a planner meant to protect must optimise against several estimators at once
and be tested on one it has never seen. The rendering layer, including whether added
content can be told apart from genuine activity, is outside this experiment.
\prov{research/counterpriv/P6.md, P6_PLAN.md; out/experiments/personal-p6-20260924-v2/REPORT.md, CELLS.csv}

\section{Lessons from building the analyser}
\label{sec:lessons}

The lessons below come from the organisation pack; the first two apply to any pack
that trains the encoder on teacher-labelled text, and the last three to anyone who
measures an inference system.

\paragraph{For rare fact types, the corpus matters more than the training.}
With company web pages alone, four of the six additive keys (stated objective,
supplier dependency, business change, hiring focus) predicted nothing, and neither
longer training nor per-key loss weighting changed that. A cue scan explained why:
3,969 web passages contained 109 candidate statements, and a focused crawl of
investor, strategy and careers pages added 48 in 2,029 passages. Firms' websites do
not say ``we rely on a single supplier'' or ``our priority this year is
profitability''; their 10-K filings do, and 2,668 filing passages from 76 firms held
107, 114 and 224 candidates for objective, dependency and business change.
Labelling 828 filing passages (cue candidates plus 400 random negatives, so that the
model does not learn to fire only on cue phrases) turned four dead keys into working
ones.

\paragraph{Language-model annotators drift; add labels, do not relabel.}
A fresh labelling run of the same passages by the same teacher, with a guide
extended only by new keys, shifted its own assertion convention: ``observed'' rose
from 28.5\% to 51.9\% of rows, where the evaluation annotator uses it for 7.7\%. A
model trained on the relabelled set fell to 0.43 assertion agreement with the
evaluation annotator, from 0.80. Labelling the corpus twice, old keys under the old
guide and new keys under the new, and merging per passage restored it. Teams that
label with language models should treat each labelling run as a separate annotator
with its own conventions.

\paragraph{Calibration data must contain the dependencies the rules rely on.}
Several weak-link rules that look right on the authored firms (a Delaware
registration indicates a corporation, 3 of 3; prices updated while credit holds
indicates holding pricing power, 7 of 10) show no lift on the generated calibration
families, because the simulator draws those variables independently. Where the
generator does encode a dependency, as it does between postings and hiring stance,
the rule's lift is measured and replicates on sealed test families. The limit is the
calibration data, not the rules; a calibration set with realistic dependencies,
which the sixteen authored firms cannot be because the system was developed on
them, is required before such rules can be promoted.

\paragraph{Matching values without relevance manufactures evidence.}
Our first backward search for evidence behind correct guesses matched a row's value
alone, and reported twenty cells whose evidence no reader used. On inspection
sixteen were false matches: the word ``growing'' in an employee's review counted as
evidence for a growth objective, and a conference venue counted as a headquarters
city. Restricting the search to rows relevant to the question reduced the twenty to
four. Readers built for the sixteen would have learned to answer from noise; the
search that finds evidence must be held to the same relevance standard as the
readers that use it.

\paragraph{Every quoted number needs a producing artefact.}
During drafting, one headline figure existed only as the output of an ad-hoc
command, two figures mixed validation and held-out results, and a precision figure
mixed certified answers with statistical ones until the certificates made the
distinction mechanical. Every number in this paper is now reproducible from a named
file or script (Appendix~\ref{app:provenance}), and the readers table is produced by
one script under one definition. We recommend the practice to anyone reporting on
inference benchmarks, where the same metric name routinely hides different
denominators.

\section{Related work}
\label{sec:related}

\paragraph{Inference from public text.}
\citet{staab2024beyond} showed that language models infer personal attributes from
users' posts, and later studied language-model anonymisers as a
defence~\cite{staab2025anonymizers}; synthetic benchmarks
followed~\cite{yukhymenko2024synthpai}. Our line of work began independently of
theirs, on individuals' own data exports, and is convergent with it. This paper
differs in treating persons and organisations as instances of one framework, in
separating reading from leakage, and in doing without a language model at analysis
time. Contextual integrity~\cite{nissenbaum2004privacy} frames why such inference is
a privacy harm: facts flow out of the context in which they were shared. We measure
that flow for a fixed question bank rather than judge its norms.

\paragraph{Quantifying what leaks.}
Quantitative information flow and $g$-leakage measure what an observation reveals
about a secret for a stated adversary~\cite{smith2009foundations,alvim2012gleakage};
Pufferfish specifies which secrets must stay
indistinguishable~\cite{kifer2014pufferfish}. The inference problem in databases
and the mosaic theory describe how individually harmless releases
combine~\cite{farkas2002inference,pozen2005mosaic}. The empirical case for the mosaic was made in 1951, when
public sources alone reproduced most of a classified order of
battle~\cite{kent1951yale}, and it is now a policy requirement for public data
releases~\cite{omb2013opendata}. Our leakage rate is an empirical counterpart for a
fixed question set: the fraction of private facts that the public record supports.

\paragraph{Information extraction.}
Span-based joint entity and relation extraction~\cite{eberts2020spert} is the
nearest family to our head-anchored rows; our design adds per-row assertion status,
subject attribution and a deterministic value grammar that makes every output
verbatim. Budget pretraining from scratch follows~\citet{geiping2023cramming}.

\paragraph{Defences.}
Obfuscation and strategic deception are established privacy
defences~\cite{brunton2015obfuscation,pawlick2019deception}. For organisations the
economics of voluntary disclosure constrain them: silence can itself be
read~\cite{grossman1981,milgrom1981}, except under endowment uncertainty or costly
disclosure~\cite{dye1985,jung1988,verrecchia1983}, and admissible signalling is a
Bayesian-persuasion problem~\cite{kamenica2011persuasion}. Minimal hitting sets
turn explanations into repairs~\cite{reiter1987diagnosis}, and the privacy funnel
frames the utility trade-off~\cite{makhdoumi2014funnel}.

\section{Limitations, next steps and ethics}
\label{sec:limits}

\paragraph{Limitations.}
The sixteen organisations are development fixtures, read many times while the
system was built; no number on them is a held-out estimate. Supported cells are not
labelled: the unsupported fraction is an upper bound from pooled readers, and the
injection protocol provides supported cells by construction, written by us with the
key in view. Extraction is scored against a strong model annotator rather than human
gold, and no standard extraction baselines (a fine-tuned BERT-base, SpERT,
zero-shot language-model extraction) are reported. Synthetic prose is regular, which
understates a learned extractor's value; real prose shows the opposite. One real
firm is analysed, for coverage only. The results on the cohorts with the expansion
articles (Table~\ref{tab:comparison}, lower block; Figures~\ref{fig:sources}
and~\ref{fig:evshare}) are development figures: four fixes to the encoder's rows
were developed on those articles, and the articles make incidental claims that
contradict the key on questions they were not written for, which counts against
every reader alike. Both instantiations are measured on development
fixtures: the personal rules were designed while inspecting the same cohorts, and its
five cohorts are versions of the same sixteen people. The personal and organisation
references use different instruments, so their comparison is directional, not
matched. Unresolved comparator
decisions are excluded, not adjudicated, and are reported next to every precision.
No uncertainty intervals are reported; every figure is a point estimate on the
stated denominator.

\paragraph{Next steps.}
The question-anchored rows (\S\ref{sec:injection}) close part of the reading gap
but not its entity part: on the second probe it reads numbers and yes/no facts
and almost no names, categories or sentences. The first next step gives its relation
rows a free-text entity object (a competitor, a customer, a city), relabelled from
the same corpus, and calibrates the row score the encoder now records but does not
use, on validation text with enough such rows. The second is a type check on
certified arguments (\S\ref{sec:realfirm}). Three measurements remain designed and
unrun. An \emph{attribution-faithfulness measurement} for
language-model readers: each question is re-asked with only the quotation the model
cited, and the share of answers that no longer follow is the model's rate of
post-hoc attribution, the counterpart of our sufficiency replay. A decoy planner that optimises against several
estimators at once and is tested on an untouched one, with the rendering layer
evaluated separately for detectability. And the
\emph{multi-carrier defence}: secrets with several independent supports, planned as
minimum hitting sets, with transfer measured on held-out injections. Beyond these,
supported-cell labels with a human-adjudicated sample would turn the upper bound of
\S\ref{sec:unsupported} into a leakage rate, and a calibration set with realistic
dependencies would let the weak-link rules of \S\ref{sec:explanation} be promoted or
retired on evidence. A symbolic layer that composes the analyser's answers into facts
no question asks for, with a proof for every derived fact and a conditional lead for
every fact one premise short, is built and will be reported separately.

\paragraph{Ethics.}
The work is defensive: it measures what a target's own public footprint reveals
and plans changes to that footprint, on the owner's machine. Two risks follow. The
analyser could be pointed at a target by a third party; it reads only public
material, which a determined adversary can already read, and deployment requires
proof of ownership of the target. Defensive interventions could mislead markets or
third parties; the ledger excludes false or misleading material statements, treats
suppression as conditional on a stated reason, and flags interventions that others
act on (a job posting that job seekers would spend effort on, for example). Real
companies appear only through their public web pages and filings, DNS records and
certificate-transparency logs, collected at one request per second with robots
exclusions obeyed, and no personal data about
individuals is inferred. The personal instantiation handles the most sensitive
material by design: it runs only on the owner's device, analyses only the owner's
own footprint, does not profile the third parties who appear in it, and uses real
data only in a consented study reviewed before it starts. Decoys add activity to the
owner's own account rather than making statements to others; the ledger classes them
as additional signals requiring the owner's sign-off, and they are never generated
about a third party or presented as a factual claim. Language-model outputs
used to label training data come from models whose licence or terms of service
permit it; outputs obtained under terms that restrict training were used for
evaluation only.

\section{Conclusion}
\label{sec:conclusion}

We presented a framework for measuring and reducing inference exposure from public
footprints that runs on the owner's own CPU with no language model at analysis
time, and whose domain knowledge sits in replaceable packs over a shared core.
Scoring inference systems against private truth measures three things at once; in
the organisation instantiation most of the score is a matter of guessing policy,
because almost half of the questions cannot be answered from the record at all and
silence is the largest source of correct answers. Separating reading accuracy from
leakage, and creating supported cells by injection, turns the benchmark into a
diagnostic. It showed that the analyser's reading was ahead of its answering,
specified the readers that closed the gap on its own templates (from 6\% to 80\% of
stated facts recovered), and then, in a pre-registered probe with a frozen analyser,
showed that this gain does not transfer: the same facts in unseen natural prose are
recovered at 16\%, without a single wrong answer. The gap is in reading natural
prose, and it is a learned extractor's to close; encoder rows anchored on the
instrument's own questions close part of it, lifting the held-out probe from 38 to
64 of 233 and a second probe of 363 facts in free-form articles from 24 to 102, and
none of the entity part. The
local analyser answers fewer questions than the strongest hosted model, and on plain
prose the models now read more of the stated facts, but its evidence is a forward
input rather than a post-hoc attachment: its certified answers are precise, cited
verbatim, replayable and computed on the owner's machine, and with the articles in
the record 70\% of its evidence-bearing answers are right on evidence that establishes
them, against 18--56\% for the language models. On the defence side, rewriting a fact's
carrier as a true, coarser statement hid every single-carrier fact from four
language-model adversaries it was not planned against, and adding material did not.
The personal instantiation ran on the same core without changing a single
organisation answer. For people the guessing term is larger still (53\% of settled
cells against 26\% for firms), evidence reading gains on public footprints but not
yet on private archives, deletions and decoys planned against one estimator of a
routine barely affect a second. A decoy planner with no language model nevertheless
cut its target estimator's correct answers from 21 to 11 of 32 without touching the
genuine record, which makes it the ingredient a later language-model renderer needs. Two questions come next: whether planning against a
transparent analyser transfers when facts have several independent supports, and
how a learned extractor adapted to informal text changes what can be read from a
person's own archive.


\appendix
\section{Provenance of quoted figures}
\label{app:provenance}

Paths are relative to the project repository (\path{research/counterbiz/countermodel}
is abbreviated \path{cm}). Every figure in Tables~\ref{tab:provenance} and~\ref{tab:provenance2} is produced
by the named file or script; no figure is transcribed from a log by hand. The
artefacts are in the project repository and will be released with the paper.

\begin{table*}[t]\small
\caption{Every quoted figure and the artefact that produces it.}
\label{tab:provenance}
\begin{tabular}{@{}>{\raggedright}p{0.40\textwidth}>{\raggedright\arraybackslash}p{0.56\textwidth}@{}}
\toprule
\rowcolor{cbzTint}
figure & artefact\\
\midrule
never-recovered questions (56), never attempted (19), recovery distribution &
\path{ai-libs/out/unanswerable_items.json}\\
readers table (Table~\ref{tab:comparison}) and Figure~\ref{fig:sources}: guess 479, analyser 512 (42 / 3 / 9 certified), evidence-only 50, models, evidence-backed precision &
\path{ai-libs/nn/cmcompare/comparison_table.py} on \path{cm/out/experiments/headcount-fix-20260925/capture} $\rightarrow$ \path{headcount-fix-20260925/comparison-table/table.json}\\
evidence audit of the models' verbatim citations (Figure~\ref{fig:sources}): citation establishes the correct answer in 35/54 Gemini, 27/90 Gemma, 42/156 Mistral, 3/98 Llama; 1,131 cells, blind judges, second-judge sample &
\path{ai-libs/nn/cmcompare/evidence_audit.py} $\rightarrow$ \path{cm/out/experiments/evidence-audit-20260925/} (\path{audit.json}, \path{AUDIT.csv}, \path{AGREEMENT.md})\\
Figure~\ref{fig:sources}, second bars (\emph{+ articles}): analyser 610 (71 certified, 91 encoder-row L2, 448 prior or L1), models 445 / 440 / 351 / 254 correct and their audit split &
analyser: \path{comparison_table.py --view injected} on \path{cm/out/experiments/precision-fixes-20260926/capture_v4h_exp} $\rightarrow$ \path{precision-fixes-20260926/table_v4h_exp_injected/table.json}, split in \path{precision-fixes-20260926/README.md}; models: \path{v4-biz-encoder-16firms-20260925/table_with_llms_injected/table.json}; audits \path{evidence-audit-exp-20260926/}, \path{evidence-audit-exp2-20260926/}\\
Figure~\ref{fig:evshare}: share of evidence-bearing answers that are right on establishing evidence (ours 69\% of 61, 70\% of 216; models 2--56\%) &
\path{cm/out/experiments/evidence-audit-ours-20260926/} (\path{audit.json}: blind audit of our 131 L1/L2 answers; \path{second_judge/AGREEMENT.md}: 128/131, $\kappa=0.95$; \path{evidence_share.json}); model audits \path{evidence-audit-20260925/}, \path{evidence-audit-exp-20260926/}, \path{evidence-audit-exp2-20260926/}\\
held-out injection probe (38 of 233, 0 wrong; frozen analyser, protocol, run notes) &
\path{cm/out/experiments/inject-fresh-20260924/} (\path{REPORT.md}, \path{RUN_NOTES.md}, \path{evaluation/})\\
question-anchored rows on the two probes (64 / 233, 6 wrong; 66 / 2 after the range fix; 102 / 363, 14 wrong) and by answer type &
\path{cm/out/experiments/v4-biz-encoder-eval-20260925/README.md}; \path{v4-biz-encoder-16firms-20260925/DETAILS_new_cohort.md}\\
precision fixes to the encoder's rows (610 with the articles; L3 71 / 7 / 14; encoder-row L2 91 / 17; the dropped cut-off) &
\path{cm/out/experiments/precision-fixes-20260926/README.md}; \path{ai-libs/nn/runs/v4_biz_encoder/calibration_v3_score.json}\\
expanded benchmark (363 facts; 24 recovered) and the headcount restriction (certified errors 9 $\rightarrow$ 3) &
\path{cm/out/experiments/expanded-20260925/REPORT.md}; \path{headcount-fix-20260925/README.md}\\
personal encoder (F1 0.52 vs rules 0.09), teacher bake-off &
\path{research/counterpriv/out/experiments/personal-encoder-eval-20260925/}; \path{personal-teacher-bakeoff-20260925/}\\
certified answers (42/45 resolved; 37 correct and none wrong at threshold 0.5), error sources &
\path{cm/out/experiments/provenance-20260924/} (\path{REPORT.md}, \path{ANSWERS.csv}); re-run with fresh encoder offsets, identical answers: \path{provenance-offsets-20260924/}\\
support bins of certified cells (40 / 4 / 5, none insufficient) &
\path{cm/out/experiments/support-check-certs-20260924/summary.json}\\
anatomy of the prior's correct answers (359 / 68 / 34 / 4; 67 of 86 with premises extracted) &
\path{cm/out/experiments/support-check-relevant-20260924/}; \path{premise-gap-20260924/REPORT.md}\\
language-model behaviour on the 359 silent cells &
\path{cm/out/experiments/premise-gap-20260924/llm_on_cells.py}\\
language-model readers &
\path{cm/out/experiments/llm-cohort-{gemma4_26b, llama31_8b, gemini38, mistral35}/}\\
\bottomrule
\end{tabular}
\end{table*}

\begin{table*}[t]\small
\caption{Every quoted figure and the artefact that produces it (continued).}
\label{tab:provenance2}
\begin{tabular}{@{}>{\raggedright}p{0.40\textwidth}>{\raggedright\arraybackslash}p{0.56\textwidth}@{}}
\toprule
\rowcolor{cbzTint}
figure & artefact\\
\midrule
injection before readers (235 pairs; 15 recovered) &
\path{cm/out/experiments/inject20-20260923T111332Z/summary.json}\\
injection after readers (188 recovered) &
\path{cm/out/experiments/inject20-readersv2-20260923T141418Z/summary.json}\\
held-out injection probe, pre-registered protocol &
\path{cm/out/experiments/inject-fresh-20260924/PROTOCOL.md}, \path{freeze.json}\\
extraction, held-out (Table~\ref{tab:extraction}) &
\path{ai-libs/out/extractor_eval/pod_v3/eval_v3_heldout_opus.json}; rules: \path{ai-libs/out/extractor_eval/heldout_opus_rules.json}\\
extraction, validation; additive-key recall &
\path{ai-libs/out/extractor_eval/pod_v3/eval_v3_val_opus.json}; \path{eval_v3_val_me.json}\\
pretraining tokens (6.35B) &
\path{ai-libs/nn/runs/pretrain_v1/final/state.json}\\
teacher agreement (57--78\%; 94.7\%) &
\path{ai-libs/slm/agreement.py} on \path{labels_heldout_{opus,codex}.jsonl}\\
rulebook statuses; steady-hiring rule on test (0.73 vs 0.47) &
\path{cm/artifacts/business_rules_v2.json}; \path{cm/out/experiments/weak-links-20260924/REPORT.md}\\
Apple coverage (3 / 6 / 10; 10 L3) and row counts; with the question-anchored rows (13; 9 L3 + 4 L2; 2 relation and 66 metric rows) &
\path{cm/out/experiments/provenance-offsets-20260924/apple/}; \path{apple-v4-20260926/} and both row sets \path{apple-both-20260926/} (\path{README.md}, \path{COMPARISON.txt})\\
Kimiya contract of the pipeline (Figure~\ref{fig:contract}) &
desktop application, \path{apps/desktop/contracts/analyze_stages.kim} (CounterBiz application repository)\\
defence on 188 secrets (Table~\ref{tab:defence}) &
\path{research/counterbiz/decoy/experiments/decoy188-20260923T213829Z/result.json}\\
personal instantiation: prior 798 / 1,515, pooled readers, cohorts &
\path{research/counterpriv/out/experiments/personal-measurement-20260924-both-final/REPORT.md}; \path{research/counterpriv/BENCHMARK.md}\\
personal extraction (22,300 records, 52,660 rows) and solvers &
\path{research/counterpriv/P1.md}, \path{P2.md}; \path{out/experiments/personal-p1-20260924-v2/}, \path{personal-p2-20260924-v2/}\\
personal evidence reading (Table~\ref{tab:personal}; 121 / 131 public attempts) &
\path{research/counterpriv/out/experiments/personal-p5-20260924/REPORT.md}\\
personal defence (32 injected cells; routines 21$\rightarrow$13, reserve 16$\rightarrow$15) &
\path{research/counterpriv/out/experiments/personal-p3-20260924/}, \path{personal-p4-20260924/}\\
personal decoy planner (226 additions per cohort; primary 21$\rightarrow$11, median 16$\rightarrow$18, routed 798) &
\path{research/counterpriv/out/experiments/personal-p6-20260924-v2/REPORT.md}, \path{CELLS.csv}; \path{P6_PLAN.md}\\
organisation parity after adding the personal pack (57,344 answers) &
\path{research/counterpriv/out/verification/organisation-parity-p5-20260924.json}\\
corpus cue counts; teacher drift &
\path{ai-libs/slm/select_thin.py}; \path{labels_train_deepseek{,_v2}.jsonl}\\
\bottomrule
\end{tabular}
\end{table*}

\end{document}